\documentclass{aa}

\usepackage[varg]{txfonts}
\usepackage{amsmath}
\usepackage{wasysym}
\usepackage{amssymb}
\usepackage{graphicx}
\usepackage{times}
\usepackage[percent]{overpic}
\usepackage{color}
\usepackage[T1]{fontenc}
\usepackage{ae,aecompl}
\usepackage{pstricks}
\usepackage{natbib,twoopt}
\usepackage[breaklinks=true]{hyperref} 
\bibpunct{(}{)}{;}{a}{}{,}

\begin{document}

\title{A new astrophysical solution to the Too Big To Fail problem}
\subtitle{Insights from the MoRIA simulations}

\author{R.~Verbeke\inst{1}
                \and
                E. Papastergis\inst{2}\thanks{NOVA postdoctoral fellow}
                \and
                A.A. Ponomareva\inst{3,2}
                \and
                S. Rathi\inst{1,4}
                \and
                S.~De Rijcke\inst{1} }
                
\institute{Astronomical Observatory, Ghent University, Krijgslaan 281, S9, 9000 Gent, Belgium \\
        \email{robbert.verbeke@ugent.be, sven.derijcke@ugent.be}
                \and
                Kapteyn Astronomical Institute, University of Groningen, Landleven 12, 9747AD Groningen, The Netherlands \\
                \email{papastergis@astro.rug.nl, ponomareva@astro.rug.nl}
                \and            
                Research School of Astronomy \& Astrophysics, Australian National University, Canberra, ACT 2611, Australia
                \and
                IIT Roorkee, Haridwar Highway, Roorkee, Uttarakhand 247667, India
                %\email{robbert.verbeke@ugent.be, sven.derijcke@ugent.be}
                }

\date{Received date; Accepted date}

\abstract
  {} %Context
  {We test whether or not realistic analysis techniques of advanced hydrodynamical simulations
  can alleviate the Too Big To Fail problem (TBTF) for late-type galaxies. TBTF states that isolated dwarf galaxy
  kinematics imply that dwarfs live in halos with lower mass than is
  expected in a $\Lambda$CDM universe.  Furthermore, we want to identify the physical mechanisms that are responsible for this observed
tension between theory and observations.} %Aims
  {We use the {\sc m}o{\sc ria} suite of dwarf galaxy simulations to investigate whether observational
  effects are involved in TBTF for late-type field dwarf
  galaxies. To this end, we create synthetic radio data cubes of the
  simulated {\sc m}o{\sc ria} galaxies and analyse their \ion{H}{i} kinematics as
  if they were real, observed galaxies. } %Methods
  {We find that for low-mass galaxies, the circular velocity profile inferred 
  from spatially resolved \ion{H}{i} kinematics often underestimates the true circular velocity profile, 
  as derived directly from the enclosed mass. Fitting the \ion{H}{i} kinematics of {\sc m}o{\sc ria} dwarfs with a theoretical halo profile 
  results in a systematic underestimate of the mass of their host halos.
  We attribute this effect to the fact that the interstellar medium of a low-mass late-type dwarf is
  continuously stirred by supernova explosions into a vertically puffed-up, turbulent state to the
  extent that the rotation velocity of the gas is simply no longer a
  good tracer of the underlying gravitational force field. If this
  holds true for real dwarf galaxies as well, it implies that they
  inhabit more massive dark matter halos than would be inferred from their kinematics,
  solving TBTF for late-type field dwarf galaxies.} %Results
  {} %Conclusion


\maketitle

\section{Introduction}

Generally considered as the current standard model for
cosmology and cosmic structure formation, $\Lambda$CDM  is a superbly successful
theory on large, super-galactic distance scales
\citep{mamon17,Rodriguez2016,Planck2016,cai14,suzuki12}. However,
towards smaller, sub-galactic scales, and especially in the regime of
dwarf galaxies, $\Lambda$CDM encounters a number of persistent problems.

One such problem is referred to as Too Big Too Fail, or TBTF, first formulated
in the context of the Local Group. Given the many factors that
suppress star formation in dwarf galaxies, such as supernova feedback
and the cosmic UV background, visible dwarf galaxies are expected to
reside in relatively scarce high-$v_\mathrm{circ}$ dark-matter
halos. This would also agree with their small observed number
density. However, most observed Milky Way satellites have circular
velocities $v_\mathrm{circ} < 30~ \mathrm{km\ s^{-1}}$, estimated from
their stellar kinematics, indicating that these satellites seem to
live in low-$v_\mathrm{circ}$ subhalos, which are too abundant in
comparison with the observed number of Milky Way satellites
\citep{boylankolchin11, boylankolchin12}. The TBTF problem is also
present for the satellite system of Andromeda \citep{tollerud14} and
for field dwarfs in the Local Group and Local Volume
\citep[e.g.][]{ferrero12, garrisonkimmel14, papastergis15}.

Several possible solutions to this problem have been suggested. For
example, if the Milky Way were to have a smaller virial mass, then it
would also host a smaller number of massive subhalos
\citep{wang12}. Another way out is to take into account the fact that
baryonic processes, such as supernova feedback, can flatten the inner
dark-matter density distribution, converting a high-$v_\mathrm{circ}$
cuspy density profile into a low-$v_\mathrm{circ}$ cored one at
constant halo mass. By fitting the mass-dependent DC14 profile
\citep{dicintio14} to the kinematical data of the Local Group dwarf
galaxies, \cite{brook15} found that dwarf galaxies inhabit more
massive halos than previously thought, thus alleviating the TBTF
problem. Other effects that help reduce dwarf galaxy circular
velocities in the context of the Local Group include tidal stripping
\citep{sawala16}.

\citet[][henceforth referred to as P16]{papastergis16} discuss the TBTF
in field dwarfs, where only
internal baryonic effects can be invoked to reduce halo circular
velocities. In their analysis, they use
abundance matching to derive the relation between the observed
\ion{H}{i} rotation velocity inferred from the galaxy 21cm emission line
profile, $W_{50}$, and the
maximum halo circular velocity $v_{h, \mathrm{max}}$ such that the
halo velocity function (VF) found in simulations \citep{sawala15}
corresponds to the observed field galaxy VF \citep{haynes11, klypin15}. 
Hereafter, we refer to this relation between $W_{50}$
and $v_{h, \mathrm{max}}$ as the P16 relation. Then, 
these authors fit NFW \citep{NFW} and DC14 profiles to the outer-most
datapoint of the rotation curves of a set of field dwarf galaxies to
infer their $v_{h, \mathrm{max}}$. This allows them to put individual
$v_{\mathrm{rot},\ion{H}{I}}-v_{h, \mathrm{max}}$ datapoints on the
inferred statistical relation. As these authors note: ``$\Lambda$CDM can be
considered successful only if the position of individual galaxies on
the $W_{50}-v_{h, \mathrm{max}}$ plane is
consistent with the relation needed to reproduce the measured VF of
galaxies.''. As it turns out, the individual galaxies are not
consistent with the expected P16 relation.

The discrepancy between these results and those from \cite{brook15}
results from the radius at which the circular velocity is
measured:~For measurements beyond the core radius ($\gtrsim 2
\mathrm{kpc}$), fitting a DC14 profile gives similar results to using a
cusped NFW profile. The TBTF problem cannot, therefore, be (fully)
explained by core creation alone \citep[see also][]{papastergis16b}.

For the present paper, we take to heart the message from
P16:~If $\Lambda$CDM is correct, then late-type field
dwarfs should have higher circular velocities than is estimated from
their \ion{H}{i} kinematics. In order to investigate such a possible
mismatch between the maximum circular velocity as inferred from gas
kinematics and its actual value, we perform \ion{H}{i} observations of
a set of simulated dwarf galaxies. In Sect. \ref{sec:simulations},
we briefly present the {\sc m}o{\sc ria} simulations and the procedure to
construct and analyse mock \ion{H}{i} data-cubes. In Sect.
\ref{sec:results}, we fit a halo profile to the outermost datapoint of
the rotation curves of the simulated galaxies and compare with the
results of P16. In Sect. \ref{sec:discussion} we
give some possible explanations for these results. Our conclusions are
presented in Sect. \ref{sec:conclusions}.

For clarity, we define the different types of velocities used
throughout this paper here:
\begin{itemize}
\item $v_\mathrm{rot, \ion{H}{i}}(R)$: the mean tangential velocity of the
  \ion{H}{i} gas at a radius $R$ from the galaxy center. This can be
  determined from observations by fitting a tilted-ring model to the
  \ion{H}{i} velocity field or the full data-cube.
\item $v_\mathrm{circ}^\mathrm{obs}(R)$: the circular velocity derived from
  the $v_\mathrm{rot, \ion{H}{i}}(R)$ profile by correcting for asymmetric drift
  (see Sect. \ref{sec:pressure}).
\item $v_h^\mathrm{true}(R)$: the ``true'' circular
  velocity profile, inferred from the total enclosed mass profile $M(R)$ as 
\begin{equation}
v_h^\mathrm{true}(R) = \sqrt{\frac{GM(R)}{R}}.
\end{equation}
\item $v_\mathrm{out, \ion{H}{i}} = v_\mathrm{circ}^\mathrm{obs}(R_\mathrm{out})$: the
  outermost value of the rotation curve.
\item $v_{h,\mathrm{max}}^\mathrm{true} = \mathrm{max}(v_h^\mathrm{true})  $ : the maximum
  circular halo velocity.
\item $v_{h,\mathrm{max}}^\mathrm{fit}$, $v_{h,\mathrm{max}}^\mathrm{NFW}$, or
  $v_{h,\mathrm{max}}^\mathrm{DC14}$:  the maximum circular velocity
  obtained by fitting an NFW or DC14 profile to
  $v_\mathrm{out, \ion{H}{i}}$. Denoted by $v_{h,\mathrm{max}}^\mathrm{fit}$ in general and
  $v_{h,\mathrm{max}}^\mathrm{NFW}$ or $v_{h,\mathrm{max}}^\mathrm{DC14}$ when the halo profile
  is specified.
\item $W_{50}$ : the full width at half maximum (FWHM) of the galactic 21cm emission line profile, corrected for inclination to an edge-on view.

\end{itemize}

All, except for $W_{50}$ , refer to a spatially resolved kinematic measurement or calculation. $W_{50}$ on the other hand is derived from the spatially unresolved \ion{H}{i} spectrum. Since $W_{50}$ does not correspond to any specific radius, it does not generally contain enough information to estimate the mass of the host halo by fitting a certain mass profile. However, $W_{50}$ measurements exist for large samples of galaxies, which allows for an accurate measurement of the number density of galaxies as a function of $W_{50}$, that is, the VF.

\section{The MoRIA simulations}
\label{sec:simulations}

We use the {\sc m}o{\sc ria} (Models of Realistic dwarfs In Action) suite of
$N$-body/SPH simulations of late-type isolated dwarf galaxies. These simulations
are the result of letting isolated proto-galaxies, starting at $z=13.5$, merge over time along a 
cosmologically motivated merger tree \citep{cloetosselaer14}. The result is 
a galaxy with a relatively well constrained halo mass at $z=0$. This approach allows
us to reach a resolution of $10^3-10^4~\mathrm{M}_\odot$ for the baryonic components and
a force resolution of $5-15~\mathrm{pc}$, without being computationally too expensive. The resolution of 
dark matter particles is scaled with the cosmic baryon fraction $f_\mathrm{bar} = 0.2115$,
so that the number of baryonic and dark matter particles is the same.

The gas can cool radiatively and be heated by the cosmic UV background \citep{derijcke13}. 
Once a gas parcel is dense enough, it is allowed to form stars. Stars inject
energy in the interstellar medium (ISM) in the form of thermal feedback by young, 
massive stars and supernovae of types {\sc i}a and type {\sc ii}. 
The ISM absorbs 70\% of the energy injected. A significant part of this energy
is used to ionise the ISM \citep{vdb13}, which further decreases
the effective energy coupling. To reduce excessive star formation at high redshift,
we take the effects of Population {\sc iii} stars into account. Stellar particles
born out of extremely low-metallicity gas ($\mathrm{[Fe/H]} < -5$) are assumed to have
a top-heavy IMF \citep{susa14}, resulting in earlier and stronger feedback \citep{heger10}.
It is important to note that the atomic hydrogen density of every gas
particle has already been computed, based on its density, temperature,
composition, and incident radiation field, to be used in the subgrid
model of the {\sc m}o{\sc ria} simulations, as described in
\citet{derijcke13}. Thus, all \ion{H}{i} observables we describe below
are directly derived from the simulations without any extra assumptions or
approximations. More details concerning the setup and subgrid physics model of these simulations
can be found in \citet[][V15]{verbeke15}, along with a demonstration of its validity. 

Since this paper, more
simulations were run with different masses and merger histories, but
the conclusions presented in V15 still stand. At the moment of writing, 
{\sc m}o{\sc ria} consists of $\sim 30$ dwarf galaxy simulations, of which we discuss 10 in more detail.
An overview of some of the basic properties of the 10 {\sc m}o{\sc ria} dwarfs discussed
in this paper is presented in Table \ref{tab:properties}. M-1 to M-5 (M-6 to M-10) have a mass resolution of $4230~\mathrm{M}_\odot$
($10515~\mathrm{M}_\odot$) for its baryonic component and a force resolution of $9.8~\mathrm{pc}$ ($13~\mathrm{pc}$).

\begin{table*}[t!]
\begin{minipage}{0.95\textwidth}
\begin{center}
\resizebox{\textwidth}{!}{
\begin{tabular}{llllllllllll}
\hline

(1) & (2) & (3) & (4) & (5) & (6) & (7) & (8) & (9) & (10) & (11) & (12)\\ 
Name &	Symbol		&	$\log_{10}(M_\star)$		&	$\log_{10}(M_\mathrm{HI})$ 	&	$\log_{10}(M_{200})$		&  $M_V$	&	$q$ &	$R_\mathrm{out}$	&	$v_{\mathrm{out}, \ion{H}{i}}$	&	$v_{h, \mathrm{max}}^\mathrm{true}$		& $W_{50}/2$ &	$\sigma_\star$	\\
		&		&	[$\mathrm{M}_\odot$]	&	[$\mathrm{M}_\odot$]	&	[$\mathrm{M}_\odot$]	 &  [$\mathrm{mag}$]	& & [$\mathrm{kpc}$]	&	[$\mathrm{km~s^{-1}}$]	&	[$\mathrm{km~s^{-1}}$]		& [$\mathrm{km~s^{-1}}$] &	[$\mathrm{km~s^{-1}}$]	\\ \hline
M-1	& $\Diamond$		& 6.56	& 7.48	& 9.97	& -12.35	& 0.49	& 1.68	& 21.66	& 36.16	& 14.30	& 10.84\\
M-2	& $\star$		& 6.59	& 7.29	& 9.95	& -12.52	& 0.63	& 1.21	& 39.87	& 36.43	& 24.35	& 18.51\\
M-3	& $\Circle$		& 6.87	& 7.31	& 9.93	& -12.77	& 0.74	& 1.13	& 19.45	& 33.81	& 15.00	& 12.95\\
M-4	& $\bigtriangleup$	& 7.41	& 7.83	& 10.00	& -13.92	& 0.41	& 2.37	& 17.04	& 37.60	& 18.58	& 12.55\\
M-5	& $\bigtriangledown$	& 7.55	& 7.34	& 10.01	& -14.35	& 0.53	& 1.09	& 41.69	& 43.10	& 41.53	& 18.89\\
M-6	& $\triangleleft$	& 7.71	& 7.94	& 10.55	& -14.93	& 0.76	& 1.93	& 39.04	& 52.72	& 27.02	& 21.60\\
M-7	& $\triangleright$	& 8.00	& 8.49	& 10.41	& -15.69	& 0.58	& 4.18	& 36.31	& 46.96	& 21.57	& 19.79\\
M-8	& $\Square$		& 8.33	& 8.64	& 10.47	& -16.50	& 0.61	& 4.87	& 45.79	& 54.49	& 29.33	& 26.49\\
M-9	& $\pentagon$		& 8.53	& 8.59	& 10.42	& -16.75	& 0.60	& 3.43	& 40.54	& 52.38	& 37.50	& 25.09\\
M-10	& $\hexagon$		& 9.07	& 8.66	& 10.84	& -18.17	& 0.56	& 4.15	& 53.94	& 67.30	& 38.27	& 33.94\\
\hline
\end{tabular}}
\caption{Properties of the 10 selected 	extsc{m}o	extsc{ria} simulations at $z=0$. (1) The name of the simulation, (2) the symbol used throughout the plots, (3) the stellar mass, (4) the \ion{H}{i} mass, (5) the halo virial mass, (6) the total V-band magnitude, (7) the intrinsic flattening of the \ion{H}{i}, (8) the \ion{H}{i} radius, (9) the outermost value of the rotation curve, (10) the maximum circular velocity of the halo, (11) the half-width-half-max of the \ion{H}{i}, and (12) the velocity dispersion of the stars at $R_\mathrm{out}$.}
\label{tab:properties}
\end{center}
\end{minipage}
\end{table*}

\subsection{\ion{H}{\sc i} disk sizes and flattening}

We aim to investigate \ion{H}{i} rotation curves, with strong focus on
the outer-most datapoint. It is therefore very important that the
simulated dwarf galaxies have realistic \ion{H}{i} disk sizes and
shapes. In V15, we already showed that the {\sc m}o{\sc ria} dwarfs have an atomic
interstellar medium (ISM) with realistic spatial substructure, as quantified
by the \ion{H}{i} power spectrum.

\begin{figure}
\includegraphics[width=0.47\textwidth]{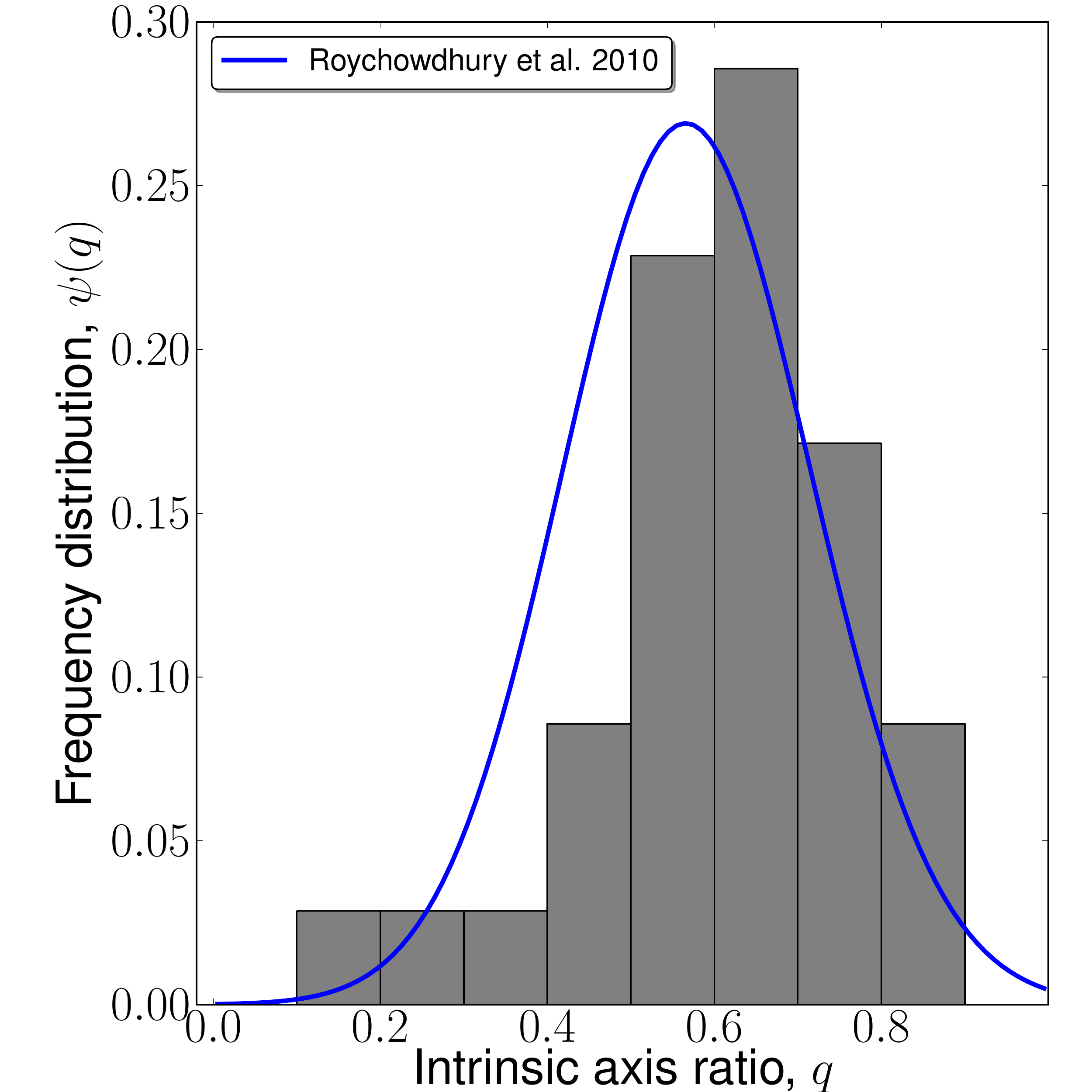}
\caption{Histogram of the axis ratios of the {\sc m}o{\sc ria} dwarf galaxies versus
  the frequency distribution obtained by \mbox{\citet{roychowdhury10}}
  for the FIGGS galaxies. \label{fig:axisratio}}
\end{figure}

\begin{figure}
\includegraphics[width=0.47\textwidth]{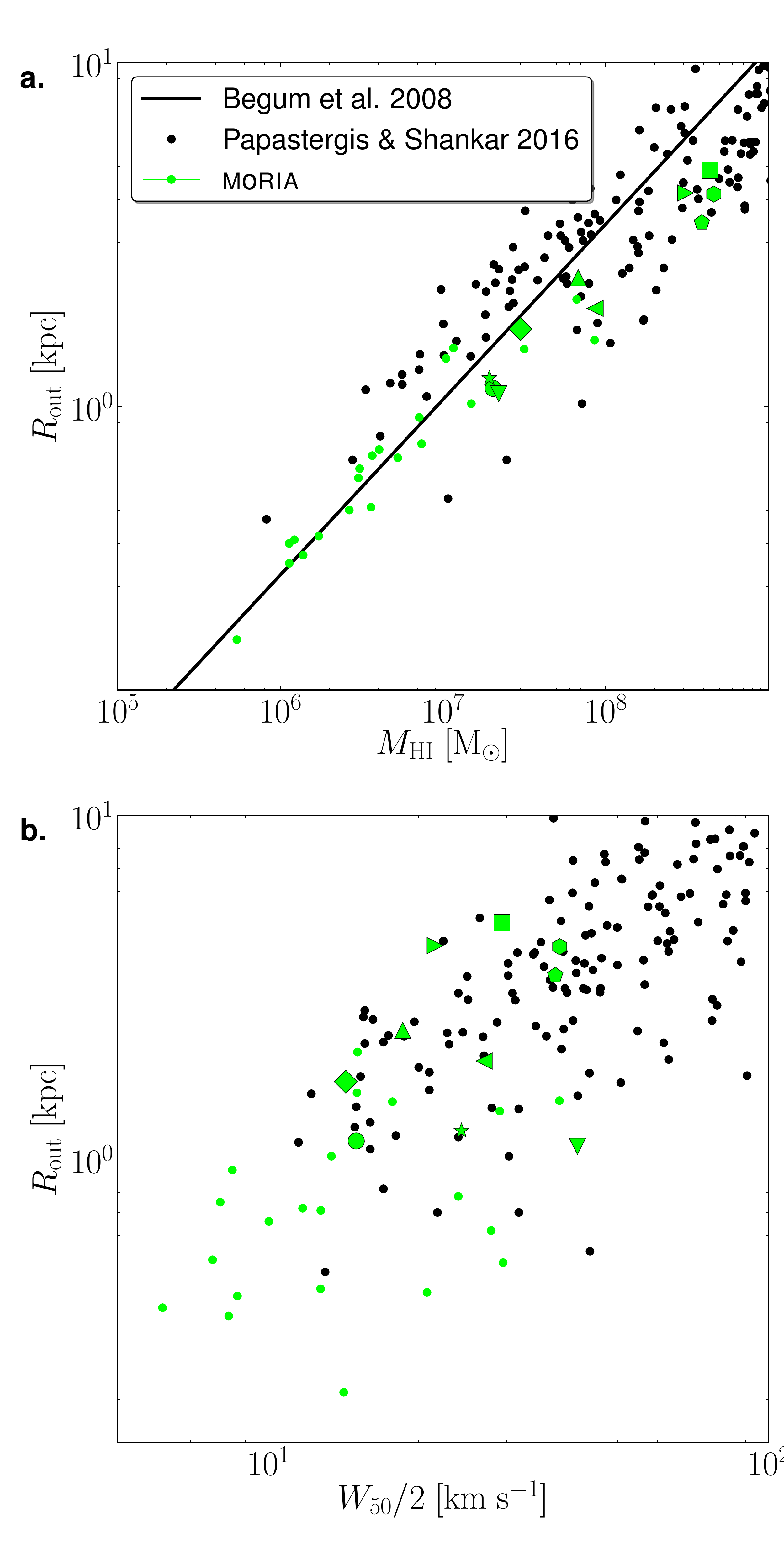}
\caption{\textbf{a.} Atomic gas mass, $M_\mathrm{\ion{H}{i}}$ versus \ion{H}{i} disk size, $R_\mathrm{out}$, and \textbf{b.} $W_{50}/2$ versus $R_\mathrm{out}$ of the {\sc m}o{\sc ria} dwarfs (green, with symbols as indicated in Table \ref{tab:properties}. Green dots indicate simulations not in the Table.)
  versus observations compiled in P16 (in black). For the simulations,
  $R_\mathrm{out}$ is the semi-major axis of the best-fitted ellipse
  to the contour with $\Sigma_\mathrm{\ion{H}{i}} =
  1\ \mathrm{M_\odot}\ \mathrm{pc}^{-2}$. The relation in panel \textbf{a.} is the one found for the FIGGS sample at $1\ \mathrm{M_\odot}\ \mathrm{pc}^{-2}$ \citep[$\log(M_\mathrm{\ion{H}{i}}) = 1.96\log(2R_\mathrm{out}) + 6.37$;][]{begum08a}. \label{fig:sizes}}
\end{figure}

Here, we also investigate the flattening and size of the \ion{H}{i}
disks. For this, we produce \ion{H}{i} surface-density contour maps of
the \ion{H}{i} and fit ellipses to the contour corresponding to a
column density of $\Sigma_\mathrm{\ion{H}{i}} =
1\ \mathrm{M_\odot}\ \mathrm{pc}^{-2} \approx 1.25 \times 10^{20}
m_\mathrm{H} \mathrm{cm}^{-2}$. We do this for different orientations
and take the minimum value of the flattening $q$, defined as the ratio
of the minor and major axis of the ellipse. In other words, $q$ is the intrinsic
axis ratio of the galaxy. The frequency distribution of the axis ratio
$q$ of the simulated {\sc m}o{\sc ria} galaxies is shown in Fig.
\ref{fig:axisratio}, along with that of observed dwarf galaxies,
derived by \cite{roychowdhury10} for the FIGSS sample of faint
galaxies. In Fig. \ref{fig:sizes}, we show the total \ion{H}{i} mass,
denoted by $M_\mathrm{\ion{H}{i}}$, as a function of the disk size,
$R_\mathrm{out}$, both for simulated and observed galaxies. The disk
size is defined as the major axis of the elliptical contour
corresponding to a column density of $\Sigma_\mathrm{\ion{H}{i}} =
1\ \mathrm{M_\odot}\ \mathrm{pc}^{-2}$.

We generally find good agreement with the observed flattening
distribution, although the {\sc m}o{\sc ria} dwarfs appear to have slightly thicker
\ion{H}{i} disks than the observed dwarfs. However, the {\sc m}o{\sc ria} dwarfs
were not intended to be equivalent to the FIGGS sample. Indeed, most
of the FIGGS galaxies have $M_\mathrm{\ion{H}{i}}\sim
10^7-10^9\mathrm{M_\odot}$ \citep[Fig. 1c in][]{begum08b} whereas more than half of the
{\sc m}o{\sc ria} dwarfs lie in the $M_\mathrm{\ion{H}{i}} \sim 10^6-10^7$
regime (see Fig. \ref{fig:sizes}). \cite{roychowdhury10} also note that galaxies with high
inclinations may be overrepresented in their sample which might lead
to a slight underestimate for the mean intrinsic axis ratio $\langle q
\rangle$. Furthermore, they assumed in their analysis that the gas
disks are oblate spheroids, and showed that $\langle q \rangle$ would
be higher when assuming a prolate spheroid. Galaxies are not
necessarily oblate spheroids \citep[e.g.][]{cloetosselaer14}, and therefore the
real $\langle q \rangle$ might be higher. Considering these points, it is
remarkable that we find a distribution that looks so similar to the
observed one.

As can be seen in Fig. \ref{fig:sizes}, the sizes of the \ion{H}{i}
disks of the {\sc m}o{\sc ria} dwarfs are also realistic;~they follow the same mass-size relation
and  FWHM-size relation as the observed galaxies compiled in P16. This is of crucial importance because it determines the
position of the outermost datapoint to which the circular velocity
profile is fitted in order to estimate $v_{h,\mathrm{max}}^\mathrm{fit}$.

\subsection{Mock data cubes}

\begin{figure*}
%%\ContinuedFloat
\begin{minipage}{\textwidth}
\begin{center}

\includegraphics[width=0.95\textwidth]{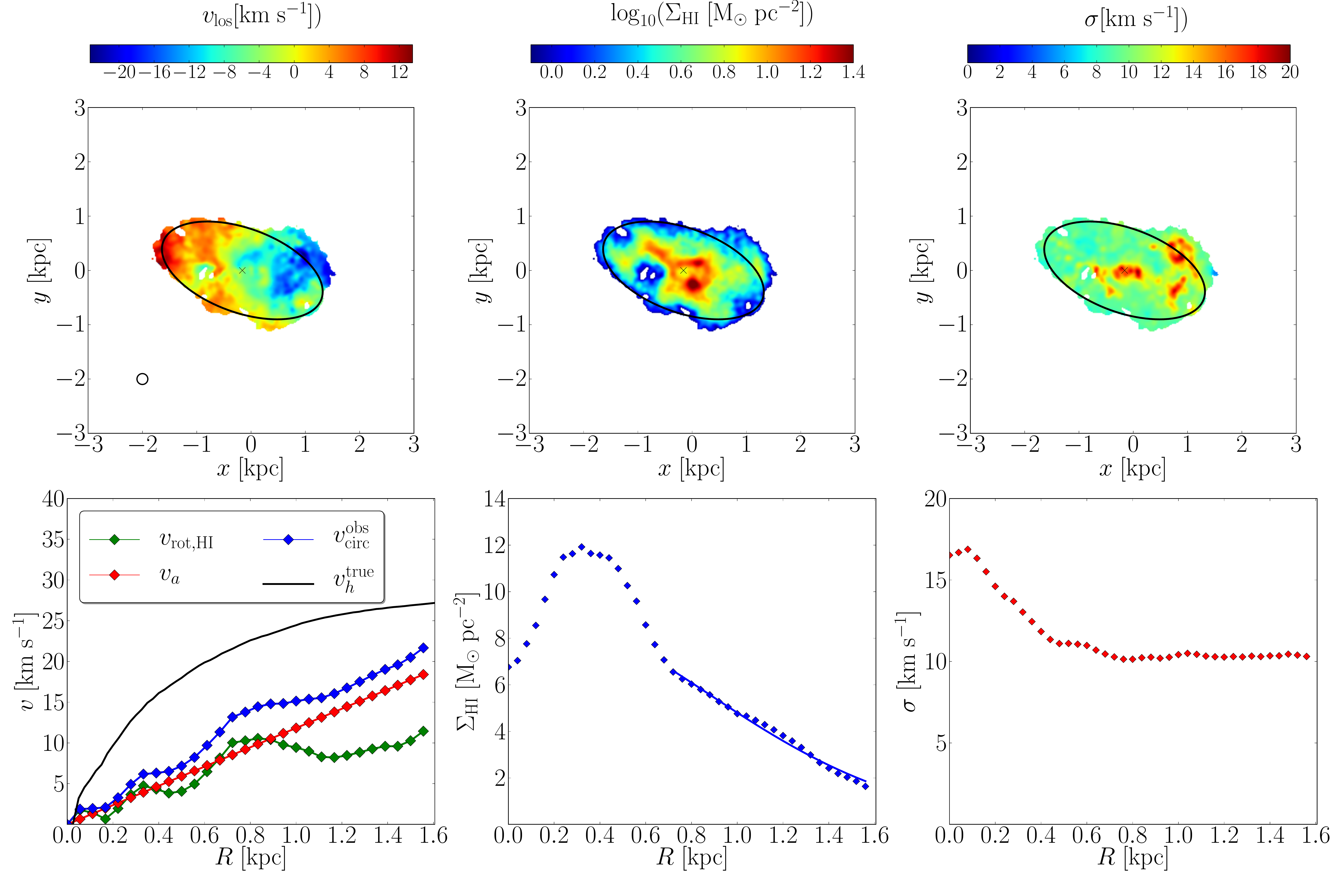}
\includegraphics[width=0.95\textwidth]{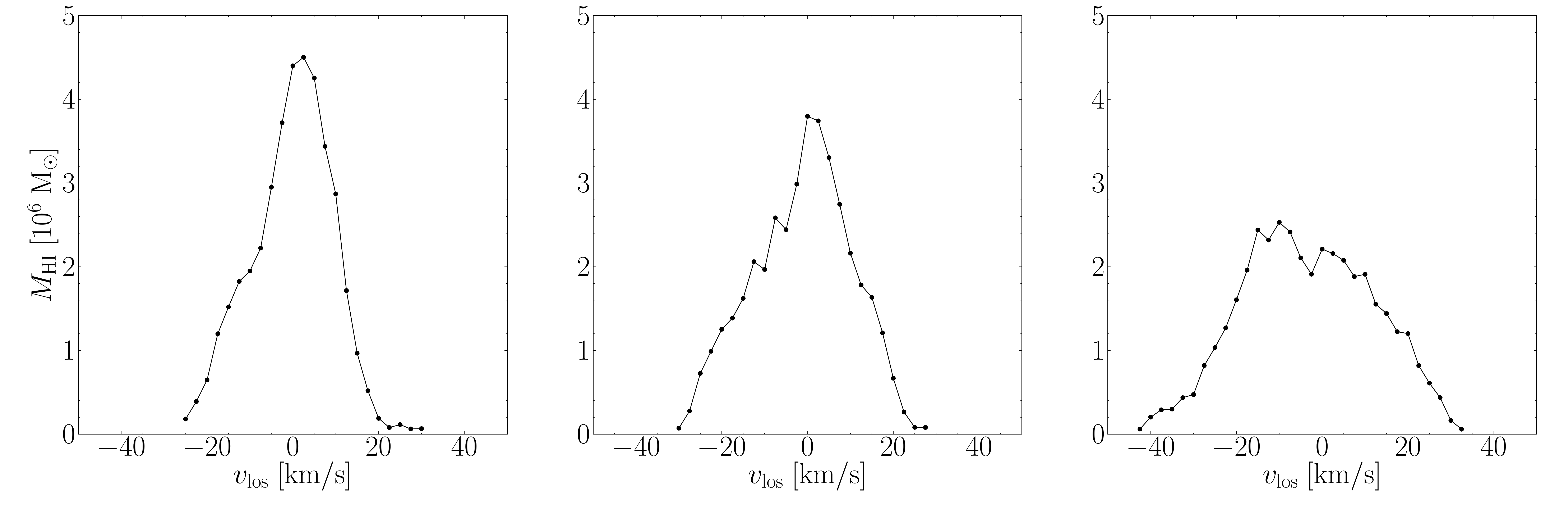}
\end{center}

\end{minipage}

\caption{\textbf{Top panels}: velocity field (\emph{left}), density
  (middle) and dispersion (right) map of M-1. The
  beam size used is shown in the top left
  panel. \textbf{Middle panels}: The rotation curves, with the
  rotational velocity obtained using tilted ring fitting in GIPSY in
  green, correction for pressure support in red, full circular
  velocity in blue and theoretical halo velocity in black (left),
  \ion{H}{i} density profile, with the fit necessary for the pressure
  support correction plotted as a solid line shown over the area that
  was used for the fit (middle), \ion{H}{i} velocity dispersion
  profile (right). \textbf{Bottom panels}: the global \ion{H}{\sc i} profile
  for the inclined view (left) as shown in the top panels and for two edge-on
  views (middle and right). The velocity bin width is the same as the channel width:
  $2.5~\mathrm{km~s^{-1}}$.
  The other {\sc m}o{\sc ria} galaxies are shown in Appendix \ref{sec:catalogue}.}

\label{fig:overview_sim}
\end{figure*}

The procedure to produce a cube of 21cm data for a {\sc m}o{\sc ria} dwarf is as
follows. First, we tilt the galaxy such that its angular momentum
vector is inclined by 45$^\circ$ with the line of sight. Then, the
mass of each gas particle is assigned to a cell in a three-dimensional
grid based on its projected position and its line-of-sight
velocity. The velocity grid is chosen with a resolution of $2.5\ \mathrm{km\ s^{-1}}$. 
To account for thermal broadening, the \ion{H}{i} mass of
each gas particle is smeared out over neighbouring velocity channels
using a Gaussian with a dispersion given by
\begin{equation}
\sigma_\mathrm{TB} = \sqrt{\frac{kT}{m_p}},
\end{equation}
where $T$ is the temperature of the particle, $k$ is the Boltzmann
constant, and $m_p$ is the proton mass. The gas is allowed to cool
down to $T=10\ \mathrm{K}$ while it becomes fully ionised around $T \sim
10^4\ \mathrm{K}$. So the thermal broadening achieves values in the
interval $0.29\ \mathrm{km\ s^{-1}} \lesssim \sigma_\mathrm{TB} \lesssim
10\ \mathrm{km\ s^{-1}}$. Finally, each velocity channel is convolved
with a Gaussian beam profile as well. The FWHM of the beams are shown in
the top-left panels in Figs. \ref{fig:overview_sim} and \ref{fig:overview_sims_appendix1}-\ref{fig:overview_sims_appendix9}. The
beam size was chosen so that it fits at least ten times within
the \ion{H}{i} radius of the galaxy. For the simulations presented
here, this comes down to 100 pc for the ones with $R_\mathrm{out} \sim
1\ \mathrm{kpc}$ (M-1, M-2, M-3, M-5, and M-6) and $200\ \mathrm{pc}$ for the larger ones
(M-4, M-7, M-8, M-9, and M-10). The resolution of the spatial grid is chosen so
that the beam size corresponds to 5 pixels. The 3D mass grid is then saved in the FITS format.

\subsection{Rotation curves}

To achieve a realistic comparison analysis, we opt for two observational 
analysis codes to derive rotation curves for the {\sc m}o{\sc ria}
dwarfs based on their radio data cubes: GIPSY \citep[The Groningen Image
  Processing SYstem;][]{vanderhulst92} and $^{\mathrm{3D}}$Barolo
\citep{diteodoro15}. GIPSY has a built-in routine, ROTCUR, which fits
a tilted-ring model to the \ion{H}{i} velocity field \citep{begeman89}. Of
the full suite of {\sc m}o{\sc ria} dwarfs, we selected ten with velocity fields and shapes
that are sufficiently relaxed to be amenable to analysis with
ROTCUR. The ones that were not selected had a very irregular \ion{H}{i} 
morphology or velocity field due to their low masses.
$^{\mathrm{3D}}$Barolo fits a model directly to the full data cube,
which makes it useful for a comparison with the GIPSY results. The
tilted-ring model in $^{\mathrm{3D}}$Barolo is populated with gas
clouds at random spatial positions. This feature makes this code very
useful for determining the kinematics of dwarf galaxies with sometimes
highly disturbed gas distributions.

Ideally, given the way we produce the data cubes, one would expect the
centre of each ring in the tilted-ring model to coincide with the
nominal galaxy centre grid, its inclination to be $45^\circ$, and its
position angle (PA) to be $90^{\circ}$. However, strongly disturbed
and warped disks can lead to tilted-ring models with the apparent ring
centres, inclinations, and PAs significantly shifted away from their
expected values. The parameters are initially estimated by fitting an ellipse to
the isodensity contour of $\Sigma_\mathrm{\ion{H}{i}} =
1\ \mathrm{M_\odot}\ \mathrm{pc}^{-2}$. These were checked and adjusted so
that, for instance, the rotation would be around the minor axis. 
The inclination is typically fixed to its true value of $45^{\circ}$ (as determined by the position of the \ion{H}{i} angular momentum vector). If the shape of the galaxy clearly implies a different inclination, we adjust it to better match this.
The chosen ellipses are shown in the top panels of 
Figs. \ref{fig:overview_sim} and \ref{fig:overview_sims_appendix1}-\ref{fig:overview_sims_appendix9}. 
The adjustment of the parameters will typically lead to a smaller maximal radius than $R_\mathrm{out}$.
Also, the isodensity contours are not perfect ellipses (the chosen rings will thus go through areas with higher densities),
leading to $\Sigma_{\ion{H}{\sc i}}(R_\mathrm{out}) > 1~\mathrm{M_\odot}\ \mathrm{pc}^{-2}$.
The systematic velocity is chosen as
roughly the value of the centre (typically close to
$0\ \mathrm{km\ s^{-1}}$). We keep these values fixed for each
radius. The rotational velocity is thus the only parameter that is
fitted. 

\subsection{Pressure support corrections}

\label{sec:pressure}
In the low-mass systems under investigation here, pressure support is
expected to be significant, entailing a sizable correction. For the latter, we follow the approach typically used in 
observational studies of dwarf kinematics \citep[e.g.][]{lelli12b}. The pressure support correction is given by
\begin{equation}
v_\mathrm{a}^2(R) = -\sigma^2\frac{\partial
  \ln(\sigma^2\Sigma_\mathrm{\ion{H}{i}})}{\partial \ln R},
\end{equation}
where $\sigma$ is the velocity dispersion and
$\Sigma_\mathrm{\ion{H}{i}}$ is the intrinsic gas surface-density. We
assume a prescription for $\Sigma_\mathrm{\ion{H}{i}}$ of the form
\begin{equation}
%f(R) = f_0\left(R_c+1\right)\left(R_c+e^{R/R_d}\right)^{-1}.
\Sigma_\mathrm{\ion{H}{i}}(R) = \Sigma_0  \exp(-R^2/2s^2),
\end{equation}
with $s$ being a radial scale length. Since we are only interested in
$v_\mathrm{out}$, the fit is performed for the outer regions and the
velocity dispersion is assumed constant at the value at the outer edge
of the galaxy. This is justified since the radial variation of
$\sigma$ is typically small. The pressure support correction then
becomes
\begin{equation}
%v_\mathrm{a}^2 = R\frac{\sigma_p^2(R)e^{R/R_d}}{R_d}\left(R_c + e^{R/R_d}\right)^{-1}.
v_\mathrm{a}^2(R) = \sigma_\mathrm{out}^2\frac{R^2}{s^2}.
\end{equation}
We also tried other commonly used prescriptions for $\Sigma_\mathrm{\ion{H}{\sc i}}$, as well as
for $\sigma^2\Sigma_\mathrm{\ion{H}{\sc i}}$ \citep[e.g.][]{oh15}. This lead to the same general results.
Given the rotation velocity $v_\mathrm{rot}$ provided by the
tilted-ring fit and the pressure support correction $v_\mathrm{a}$, we
obtain an estimate for the true circular velocity as
\begin{equation}
v_\mathrm{circ}^\mathrm{obs}(R) = \sqrt{ v_\mathrm{rot, \ion{H}{i}}^2(R) + v_\mathrm{a}^2(R)}.
\end{equation}

The obtained rotation curves are shown in the middle left panels of
Figs. \ref{fig:overview_sim} and \ref{fig:overview_sims_appendix1}-\ref{fig:overview_sims_appendix9}. In these Figs. we show the
three first moment maps of the data cube, the rotation and circular
velocity curves, the radial \ion{H}{i} density profile, and the
\ion{H}{i} velocity dispersion profile. In the density profile
diagram, the region where the parameters of the pressure support model
have been determined is indicated. In this region, the pressure support 
correction can be computed relatively reliably; at smaller radii, the
resulting pressure support correction may not be as reliable.

\begin{figure}
\includegraphics[width=0.47\textwidth]{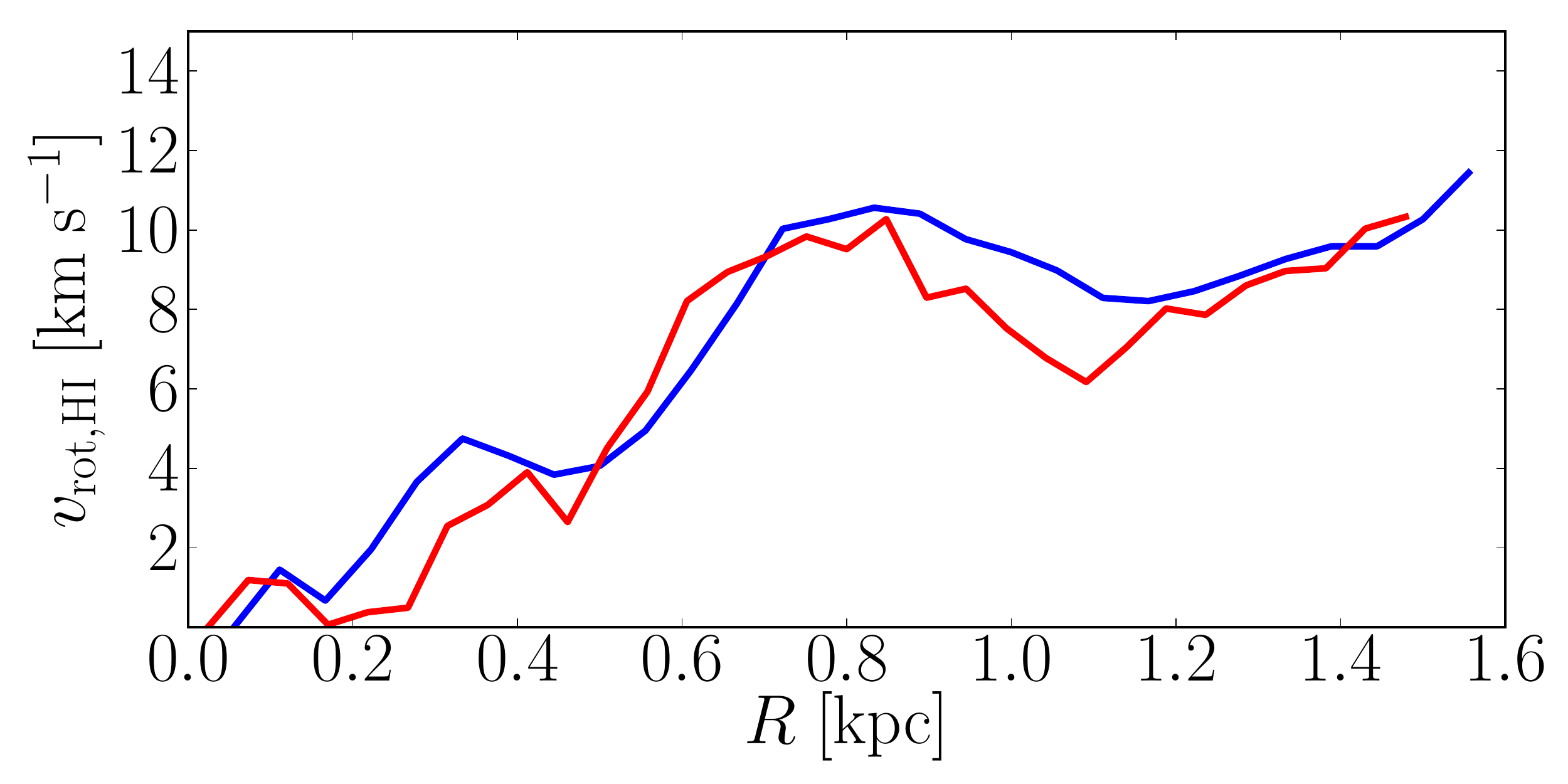}
\caption{Rotation curves of M-1, obtained using GIPSY
  (in blue) and using $^{\mathrm{3D}}$Barolo (in red). \label{fig:barolo}}
\end{figure}

\begin{figure}
\includegraphics[width=0.47\textwidth]{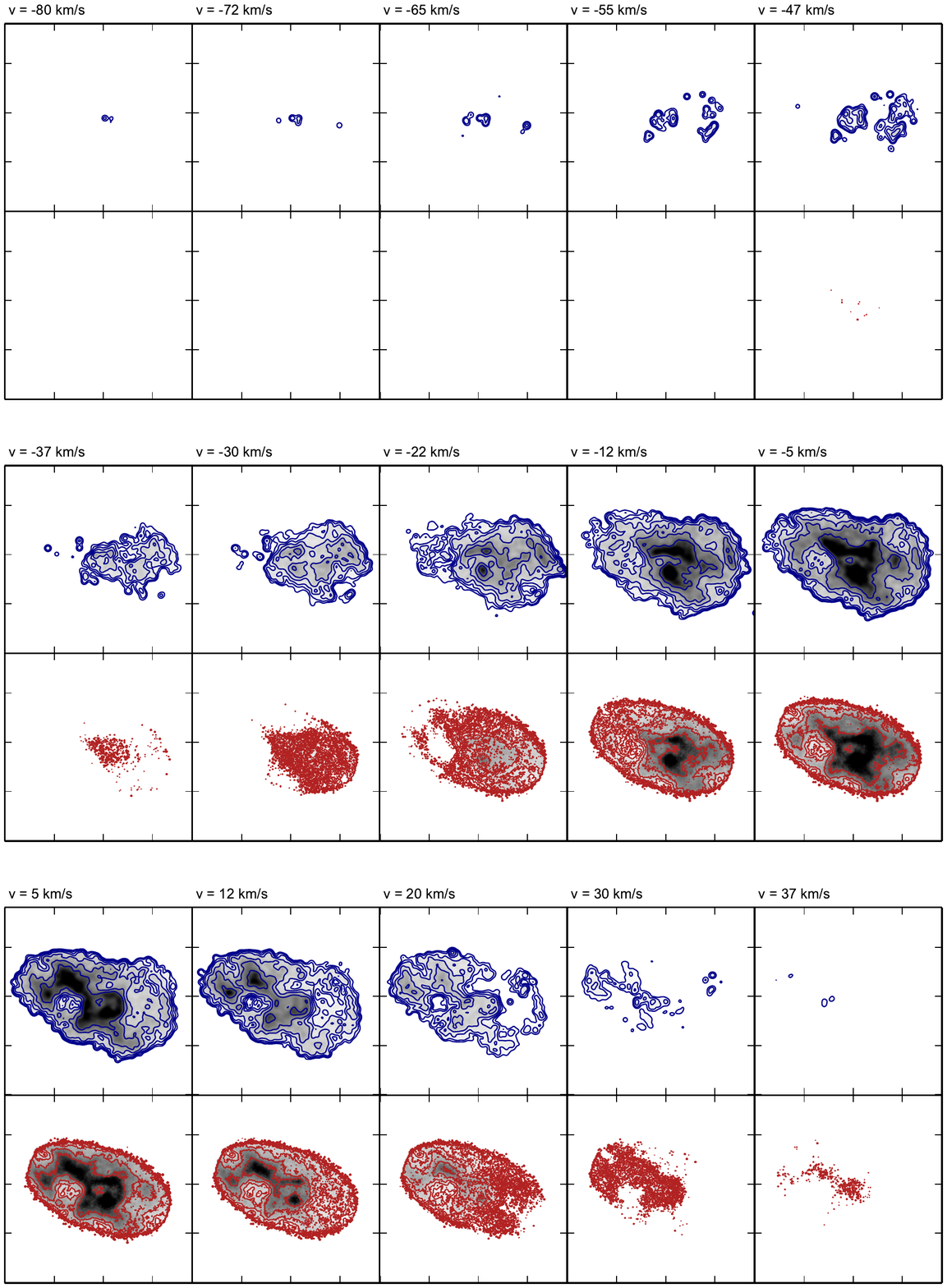}
\vspace*{-3em}
\caption{
Channel maps of M-1, in blue, and the model fit
  with $^{\mathrm{3D}}$Barolo, in red. The $^{\mathrm{3D}}$Barolo model
  reproduces the most salient features of the input data
  cubes. \label{fig:chanmapbarolo}}
\end{figure}

In Fig. \ref{fig:barolo}, we compare rotation curves determined with
GIPSY and with $^{\mathrm{3D}}$Barolo. For the tilted-ring analysis
with $^{\mathrm{3D}}$Barolo, we used 20 rings, each with a radial size
of $10~\mathrm{arcsec} \approx 50~\mathrm{pc}$. We keep all parameters
the same as in the analysis with GIPSY while fitting the
rotational velocity, with the exception of an
additional free parameter of scale height. We notice from the channel maps, two of which are
shown in Fig. \ref{fig:chanmapbarolo}, that our observed simulations
(in blue) and the model (in red) agree relatively well. Overall, the
agreement between the results obtained with both codes is
satisfactory, especially in the outer regions that are of greatest
interest to us.

We note that, since we have focused on obtaining the rotation curves and 
pressure support corrections in the outer regions, we do not make 
strong claims about the rotation in the central regions. To investigate, for example, the universality of dwarf galaxy rotation curves \citep{karukes17}
or the radial acceleration relation \citep{mcgaugh16b, lelli17}
for the {\sc m}o{\sc ria} galaxies, we would need to realistically obtain 
rotation curves at all radii. This lies beyond the scope of this paper.

\section{Results}
\label{sec:results}

\subsection{The $W_{50}-v_{\mathrm{out}, \ion{H}{\sc i}}$ relation}

\begin{figure}
\includegraphics[width=0.47\textwidth]{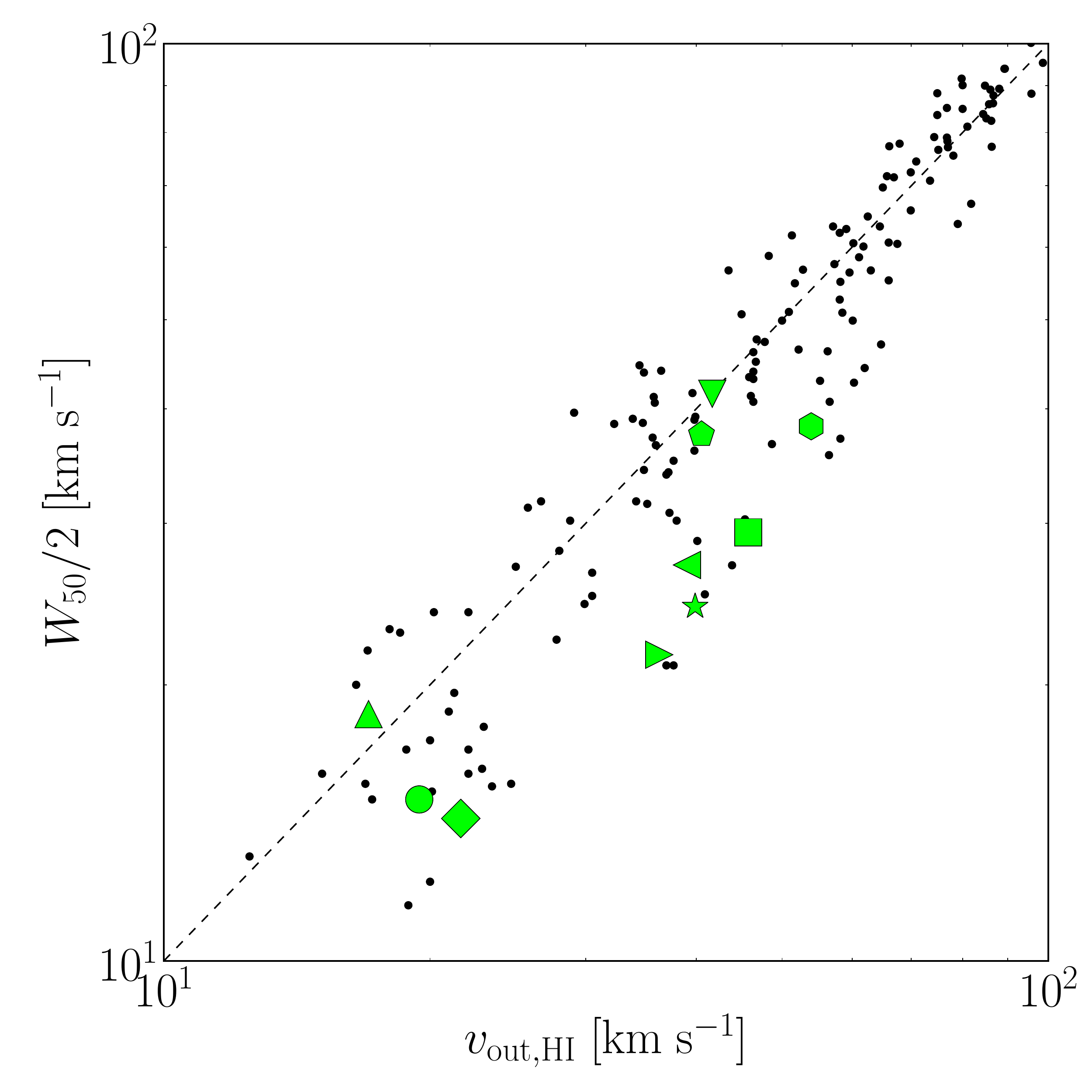}
\caption{$W_{50}/2$ versus $v_{\mathrm{out}, \ion{H}{\sc i}}$ for the {\sc m}o{\sc ria} dwarfs.
        Symbols are as indicated in Table \ref{tab:properties}. Observations, compiled in P16, are shown in black. The dashed lines show the case for both quantities being equal.  \label{fig:w50_vout}}
\end{figure}

An often-posed question is how $W_{50}$, which is relatively easy to measure, relates to the harder-to-obtain $v_{\mathrm{out}, \ion{H}{\sc i}}$ \citep[see e.g.][]{brook16, ponomareva16}. Fig. \ref{fig:w50_vout} shows the relation between the two quantities for observed low-mass galaxies (compiled in P16). For galaxies with $v_{\mathrm{out}, \ion{H}{\sc i}} \lesssim 70~\mathrm{km~s^{-1}}$, the scatter on the relation becomes significant, and it typically holds that $W_{50}/2 < v_{\mathrm{out}, \ion{H}{\sc i}}$. The {\sc m}o{\sc ria} galaxies follow the trend of the observations, although some seem to be on the low-end of the data. 

\subsection{The $W_{50}-v_{h, \mathrm{max}}$ relation}

In their study of the TBTF problem in field dwarfs, P16 derived the average 
relation between the observed \ion{H}{i} velocity width of
galaxies, $W_{50}$, and the maximum circular velocity of their host halos, $v_{h,\mathrm{max}}^\mathrm{true}$, such that the observed VF of galaxies \citep{haynes11,klypin15} can be reproduced within the $\Lambda$CDM cosmological model. The observed rotation
velocity we use here is $W_{50}/2$.

\begin{figure}
\includegraphics[width=0.47\textwidth]{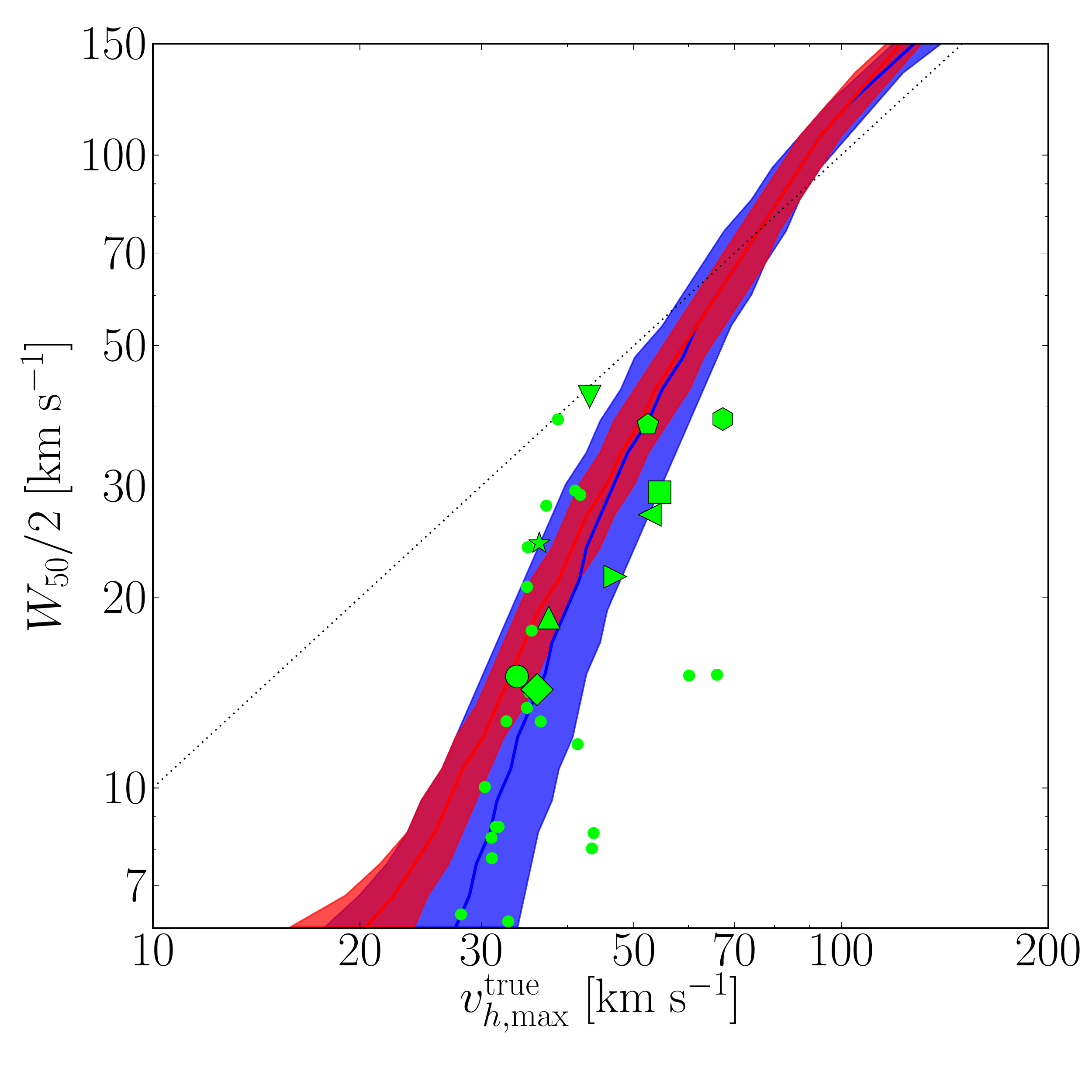}
\caption{The {\sc m}o{\sc ria} dwarfs in the $W_{50}-v_{h, \mathrm{max}}$
  plane. Green symbols are the simulations for which resolved
  rotation curves are available, with their symbols as indicated in Table \ref{tab:properties}. Green dots indicate \textsc{m}o\textsc{ria} simulations not explicitly discussed. The red and blue lines are the
  P16 relations derived from different observational datasets, with
  the bands around them representing their uncertainty. \label{fig:tbtf_w50}}
\end{figure}

Figure \ref{fig:tbtf_w50} shows the location of the {\sc m}o{\sc ria} dwarfs in the
$W_{50}-v_{h, \mathrm{max}}^\mathrm{true}$ plane. The {\sc m}o{\sc ria} dwarfs follow the average relation derived in P16 very well, a fact that ensures that {\sc m}o{\sc ria} dwarfs are produced at the correct number densities as a function of their
$W_{50}$ (i.e. the {\sc m}o{\sc ria} simulation reproduces the observational VF). Similar results were also
obtained by \citet{maccio16} based on the NIHAO hydrodynamical simulations \citep{wang15} and by \citet{brooks17} based on a set of hydrodynamic simulations of galaxy formation carried out by \citet{governato12}, \citet{brooks14} and \citet{christensen14}.

However, reproducing the observational VF alone
does not necessarily mean that the cosmological problems faced by $\Lambda$CDM on small scales have been
resolved. In particular, a successful simulation must also be able to reproduce the internal, 
spatially resolved kinematics of observed dwarfs. This is a crucial point, since the inconsistency between the
predicted velocity profiles of simulations that are able to reproduce the observational VF, and the
measured outermost-point rotational velocities of small dwarfs is at the heart of the TBTF problem (e.g. \citealt{papastergis16b}, see also \citealt{trujillogomez16, schneider16}).

\subsection{Halo profile fitting}

The NFW profile \citep{NFW} has the form
\begin{equation}
\rho_\mathrm{NFW}(R) = \frac{\rho_s}{\left(\frac{R}{R_s}\right)
  \left(1+\frac{R}{R_s}\right)^2}
\label{eq:nfw}
,\end{equation} 
and the DC14 profile is given by the expression \citep{dicintio14}
\begin{equation}
\rho_\mathrm{DC14}(R) =
\frac{\rho_s}{\left(\frac{R}{R_s}\right)^\gamma
  \left(1+\left(\frac{R}{R_s}\right)^\alpha\right)^{(\beta-\gamma)/\alpha}},
\label{eq:dc14}
\end{equation} 
where $R_s$ is scale length and $\rho_s$ is a multiple of the density
at radius $R=R_S$. The NFW profile was derived from dark-matter-only
simulations while the DC14 profile takes the halo response to baryonic
effects into account. The $\alpha$, $\beta$, and $\gamma$ parameters
are set by the star formation efficiency of the galaxy (quantified by
the ratio of stellar to halo mass, $M_\star/M_h$). For $\alpha=1$,
$\beta=3$, and $\gamma=1$, the DC14 profile coincides with the NFW
profile. If the stellar mass of the galaxy is known, both profiles have only two free parameters:~the halo mass
$M_h$ and the halo concentration $c = R_\mathrm{vir}/R_s$, with
$R_\mathrm{vir}$ , the virial radius.

P16 fitted both these profiles to the velocity measured at the outermost
\ion{H}{i} point of each galaxy (data taken from \cite{begum08a,
  deblok08, oh11, swaters09, swaters11, trachternach09, kirby12,
  cote00, verheijen01, sanders96, hunter12, oh15, cannon11,
  giovanelli13, bernsteincooper14}). They fixed the halo concentration
to the mean cosmic value \citep[$\log_{10} c = 0.905 - 0.101\log_{10}(M_h/(10^{12}h^{-1}\mathrm{M_\odot}))$;][]{dutton14}, 
leaving only the halo mass as a free parameter. 
From the fitted profile, they compute the maximum circular velocity of
each galaxy’s host halo.
In a successful cosmological model, individual galaxies should have
$W_{50}-v_{h, \mathrm{max}}^\mathrm{fit}$ data-points that agree with the average
$W_{50}-v_{h, \mathrm{max}}^\mathrm{true}$ relation that is needed to reproduce
the observed VF (blue and red bands in Fig. \ref{fig:tbtf_w50}).
As shown by P16,
all is not well;~a sizable fraction of low-mass galaxies fall to the
left of the expected $W_{50}-v_{h, \mathrm{max}}^\mathrm{true}$ relation. In other
words,~the halo circular velocity implied by their \ion{H}{i}
kinematics is too low. 

\begin{figure}
\includegraphics[width=0.47\textwidth]{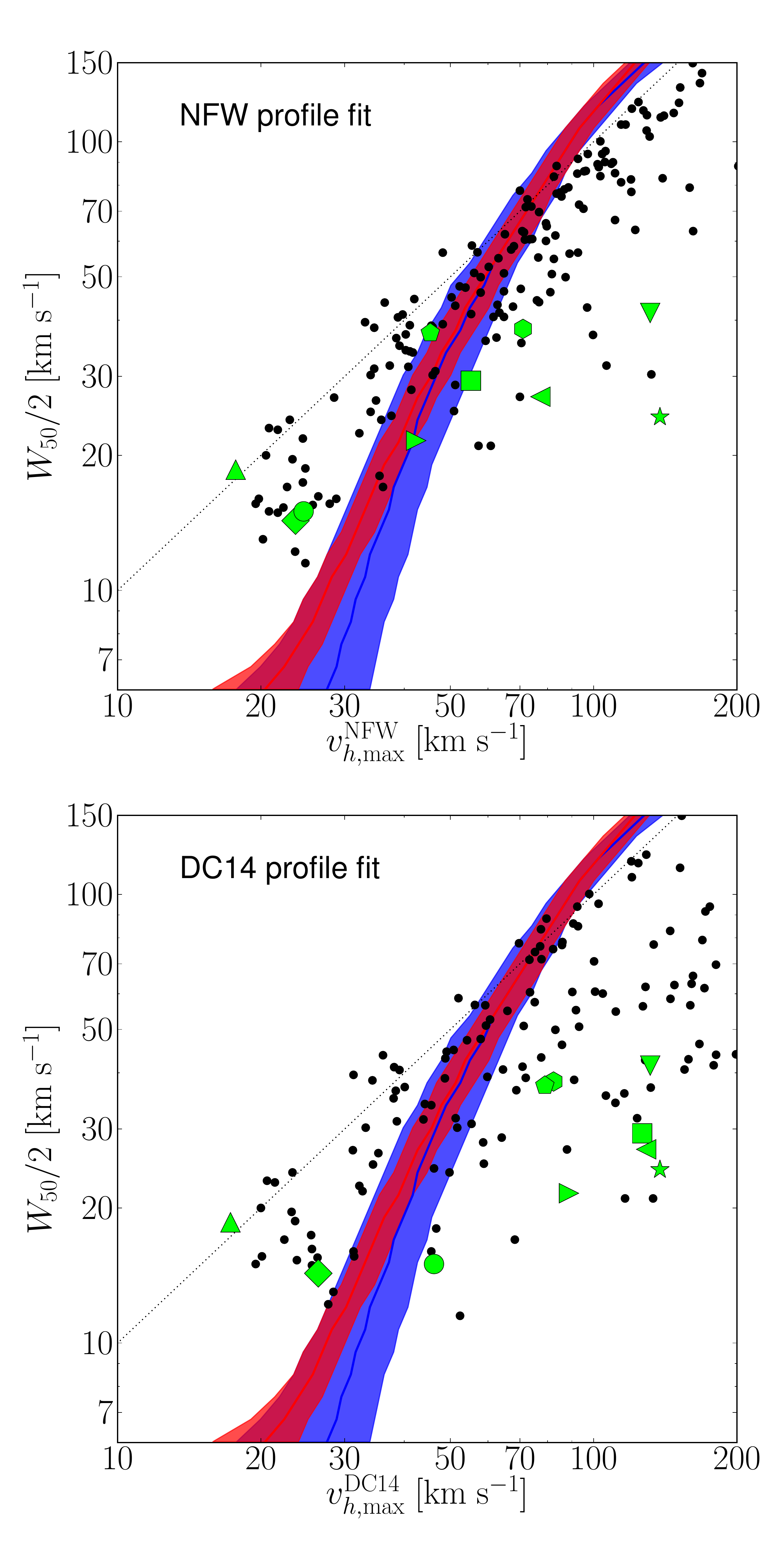}
\caption{Results from fitting a NFW (top panel) and DC14 (bottom
  panel) to the outer-most point of the rotation curves of the {\sc m}o{\sc ria}
  simulations using a fixed halo concentration (symbols as indicated in Table \ref{tab:properties}). This is compared to
  the results from P16 (in black). Red and blue
  lines and bands are the same as in Fig.
  \ref{fig:tbtf_w50}. \label{fig:halofit}}
\end{figure}

We exactly replicate this analysis for the {\sc m}o{\sc ria} dwarfs and show the
results in Fig. \ref{fig:halofit}. Although there are fewer {\sc m}o{\sc ria}
dwarfs than in the P16 sample, the result is broadly the same:~the
$W_{50}-v_{h, \mathrm{max}}^\mathrm{fit}$ relation is inconsistent with the
expected $W_{50}-v_{h, \mathrm{max}}^\mathrm{true}$ relation. The simulations
with the highest $v_{h, \mathrm{max}}^\mathrm{fit}$-values seem to lie on the low end
of, or even slightly below, the datapoints. This can be attributed to their lower-than-average
$W_{50}$-values (see Fig. \ref{fig:w50_vout}). Another explanation is their smaller-than-average
$R_\mathrm{out}$ values (see Fig. \ref{fig:sizes}a), since for smaller radii, the uncertainties
on $v_\mathrm{out, \ion{H}{\sc i}}$ will be extrapolated to large uncertainties on
$v_{h, \mathrm{max}}^\mathrm{fit}$. The two (low $W_{50}$ and small $R_\mathrm{out}$ values) most
likely work together and probably come hand-in-hand. Indeed, for smaller \ion{H}{\sc i} bodies,
the potential will be traced at smaller radii, resulting in a lower $W_{50}$.

Analysed in this way,
one would be driven to the conclusion that the {\sc m}o{\sc ria} dwarfs do not
follow the $W_{50}-v_{h, \mathrm{max}}^\mathrm{true}$ relation required for the
$\Lambda$CDM halo VF to match the observed galactic VF and, therefore, that
they suffer from the TBTF problem. The crucial
difference between Figs. \ref{fig:tbtf_w50} and  \ref{fig:halofit} is the fact that in the former, the maximum halo velocity,
$v_{h,\mathrm{max}}^\mathrm{true}$, is computed directly from the enclosed mass profile of each simulated galaxy, while in the latter,  $v_{h,\mathrm{max}}^\mathrm{fit}$ is computed by fitting the mock \ion{H}{i} kinematics of each simulated galaxy.

\section{Discussion}
\label{sec:discussion}

\subsection{Does the \ion{H}{\sc i} rotation curve trace the potential?}

\begin{figure}
\includegraphics[width=0.47\textwidth]{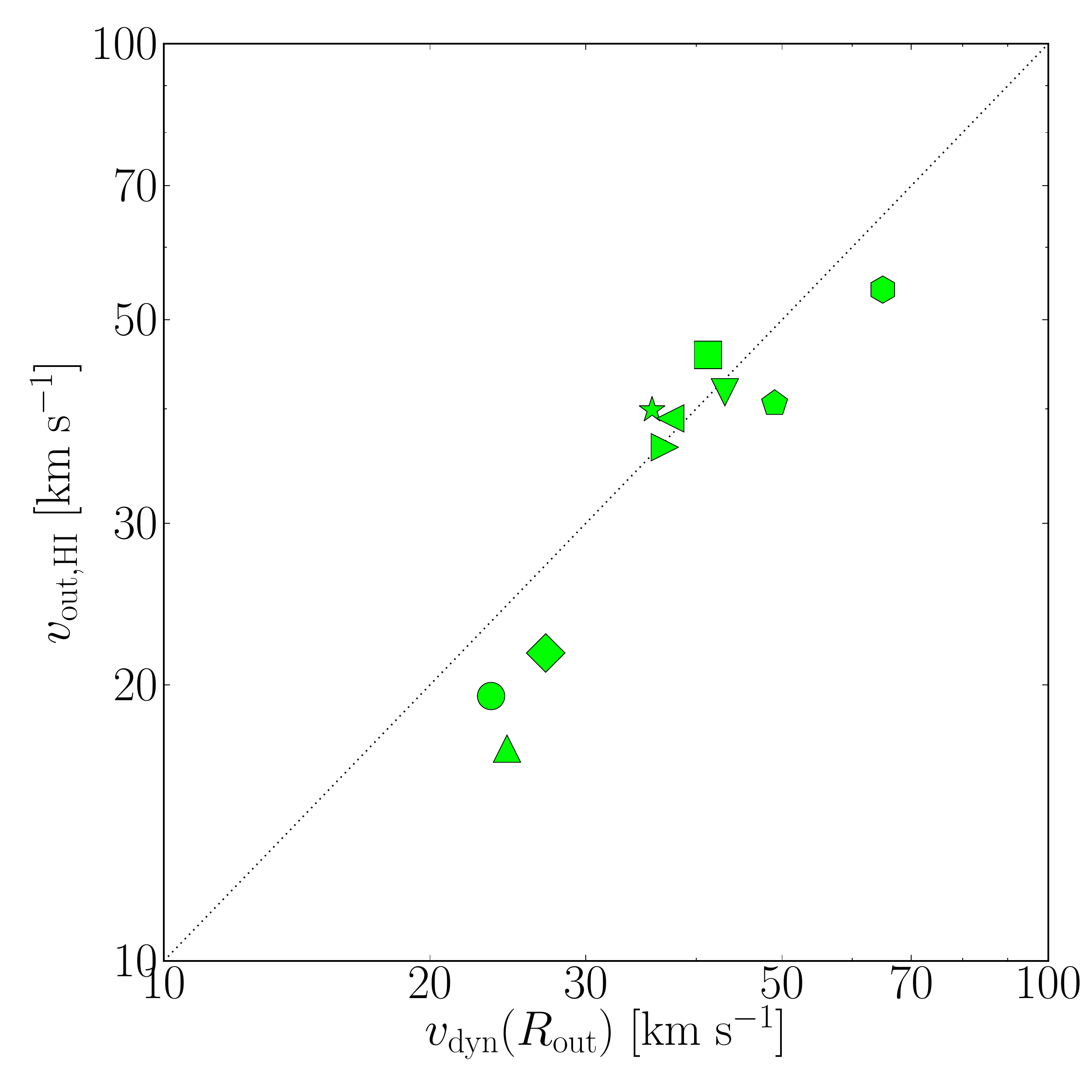}
\caption{The measured \ion{H}{\sc i} rotation at $R_\mathrm{out}$ compared to the dynamical circular velocity, as expected from the enclosed mass. 
        Symbols are as indicated in Table \ref{tab:properties}. \label{fig:expected_vc}}
\end{figure}

The fact that $v_{h,\mathrm{max}}^\mathrm{true}$ and $v_{h,\mathrm{max}}^\mathrm{fit}$
have different values might be explained by an observational effect:~The \ion{H}{i}
rotation curve of dwarf galaxies does not exactly follow the
underlying potential.  

Indeed, as one can see in Fig. \ref{fig:expected_vc}, the
\ion{H}{i} circular velocity profiles ($v_\mathrm{circ}^\mathrm{obs}(R)$) of the
{\sc m}o{\sc ria} dwarfs are quite different from the true circular velocity profiles ($v_h^\mathrm{true}(R)$), even after correcting for pressure
support. 
More often than not, {the outer \ion{H}{i} rotation velocity data-point falls
significantly below the true value of the local circular velocity}. It is important to keep in mind that the
preceding statement is not directly related to the fact that rotational velocities derived from the
linewidth of the HI profile of dwarf galaxies, $W_{50}/2$, underestimate the maximum circular velocity of
the host halo, $v_{h,\mathrm{max}}^\mathrm{true}$, a result that has already been reported by \citet{maccio16} and \citet{brooks17}. In fact, the linewidth-derived \ion{H}{i} velocity probes radii much smaller than the radius where the host halo rotation curves peak, and thus there is no guarantee that the two quantities should be the same. Further more, this is different from the fact that the \ion{H}{\sc i}
rotation curve is still rising at its outer-most radius and thus does not trace $v_{h,\mathrm{max}}^\mathrm{true}$ \citep{brook15b, ponomareva16}.
What we demonstrate here instead is that the circular velocity computed from spatially resolved \ion{H}{i} data
underestimates the true circular velocity {at the same radius}.
Of course, there are only ten {\sc m}o{\sc ria} dwarf galaxies with resolved rotation curves
and a bigger sample of simulated dwarfs is definitely required to fully
explore this issue. But if this explanation holds water, it would
explain why the halo fitting using a fixed concentration
fails;~we are not fitting to the actual halo velocity at this
radius. 

In Appendix \ref{sec:concentrationfitting}, we briefly redo the halo fitting,
but now fixing the halo mass using an abundance-matching relation and keeping the 
halo concentration as a free parameter. In short, the resulting concentrations
do not seem to be drawn from the distribution predicted by $\Lambda$CDM, especially for 
galaxies with low $W_{50}$. If the observed \ion{H}{i} rotation curve does not trace 
the potential, this would explain the seemingly incorrect population of concentrations.

\citet{valenzuela07} and \citet{pineda17} have also applied a tilted-ring method to derive the \ion{H}{\sc i}
rotation curve to investigate whether dwarf galaxies have dark matter cusps or cores. 
They studied galaxies with an idealised set-up and both find that the observed \ion{H}{\sc i} rotation underestimates the gravitational 
potential; only in the inner regions,  however.
Here, using more realistic dwarf galaxies,
we show that the idea that the \ion{H}{i} rotation is not necessarily a good 
tracer for the underlying gravitational potential of dwarf galaxies is not necessarily 
confined to the inner regions of galaxies, but extends over their entire body.

One crucial question here is what causes this substantial underestimate of the local circular velocity in
observational measurements of the HI kinematics. We attribute this to 
the fact that the assumptions underlying the
tilted-ring fitting method and the correction for pressure support are not
met in the case of low-mass dwarf galaxies;~their atomic ISM simply
does not form a relatively flat, dynamically cold disc. Rather, they
have a vertically thick ($\langle q \rangle \sim 0.5$), dynamically
hot, continuously stirred atomic ISM with significant substructures,
that is not in dynamical equilibrium in the gravitational potential.
The detailed analysis of
the vertical structure of the HI disks of  {\sc m}o{\sc ria} dwarfs and of non-circular motions in their velocity fields
will be the focus of a separate publication (Verbeke et al. in prep.)

\subsection{Stellar kinematics}

\begin{figure}
\includegraphics[width=0.47\textwidth]{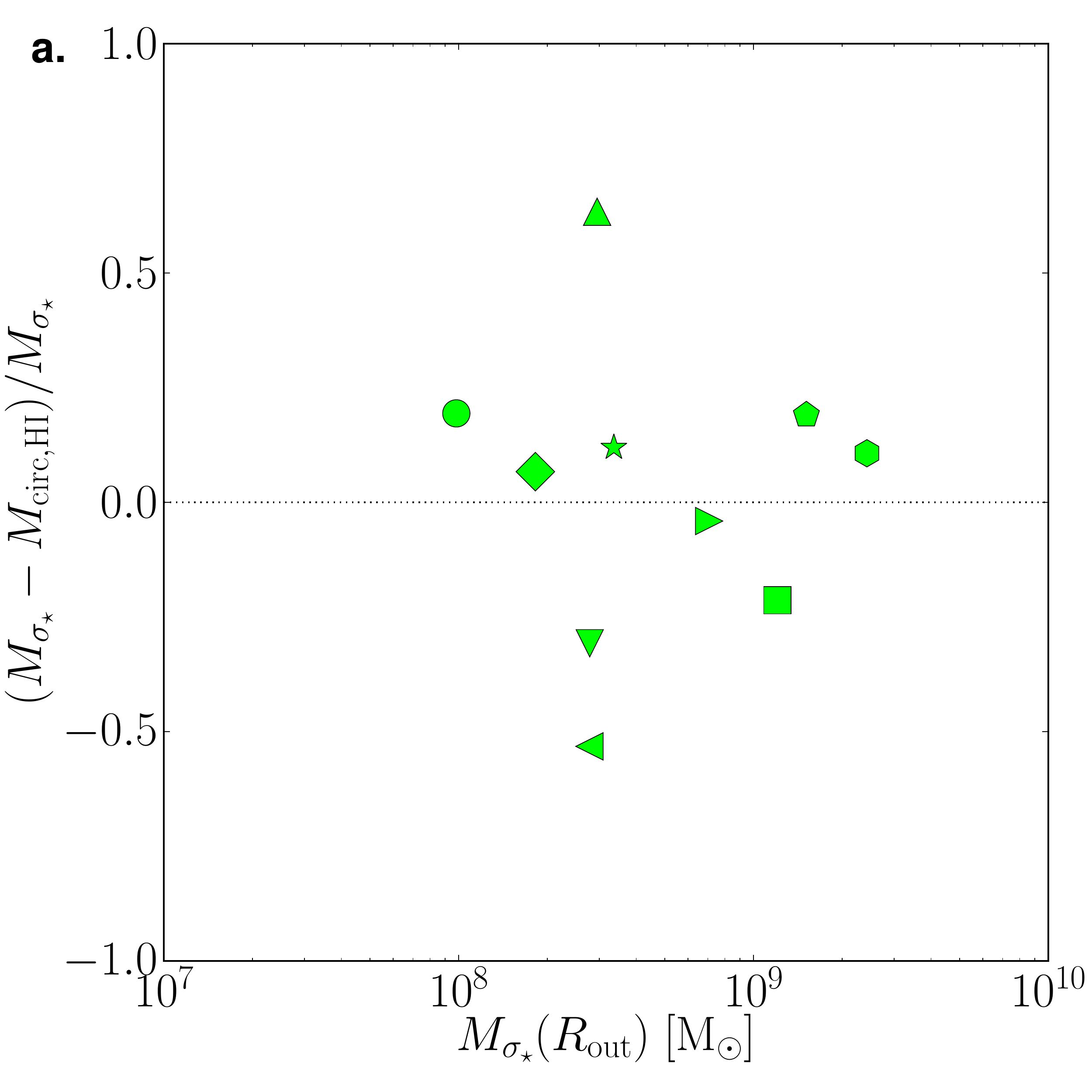}
\includegraphics[width=0.47\textwidth]{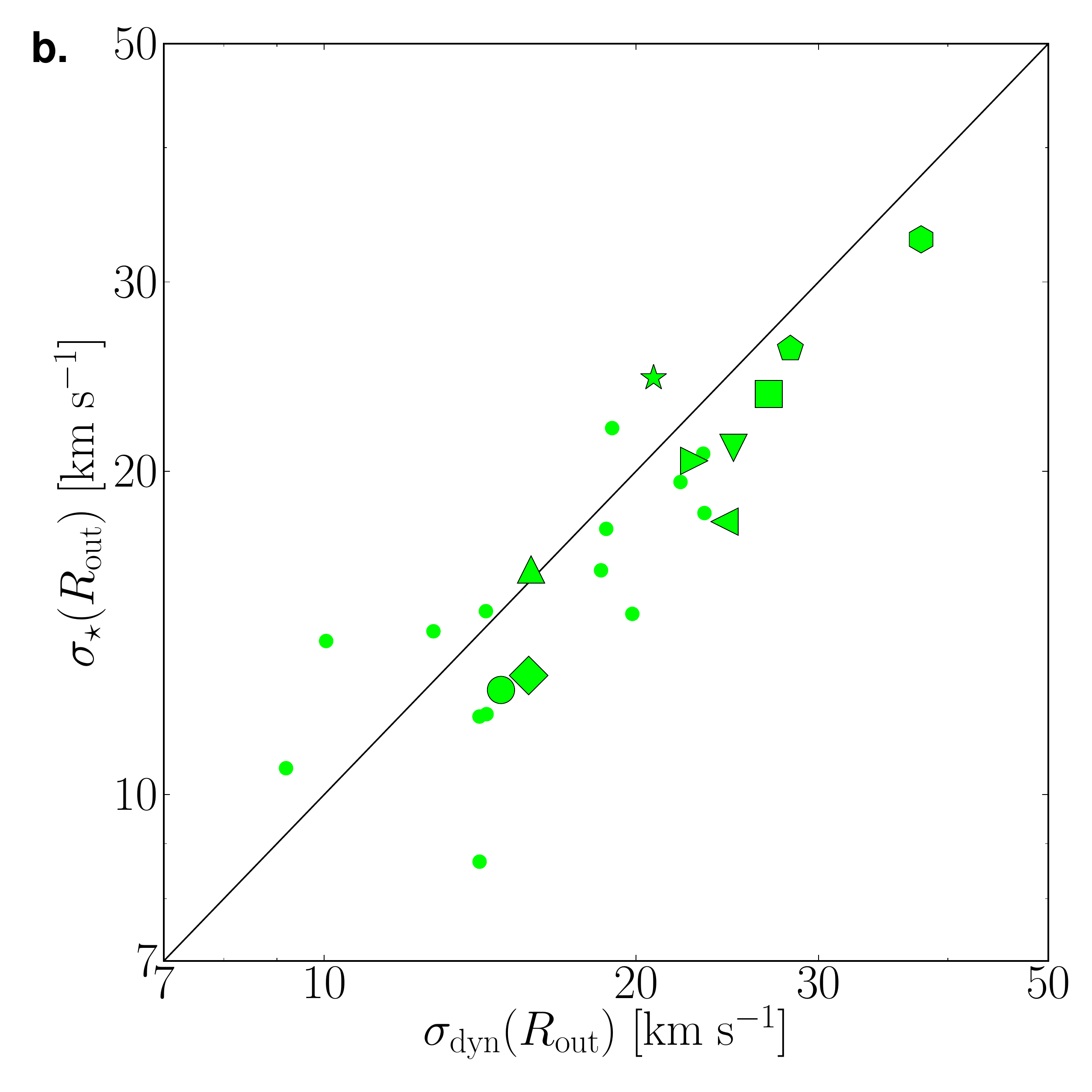}
\caption{\textbf{a.} The relative difference between the enclosed mass within $R_\mathrm{out}$ inferred from stellar velocity dispersions $\sigma_\star$ and \ion{H}{i} circular motions. \textbf{b.} The stellar velocity dispersion within $R_\mathrm{out}$ compared to the dynamical velocity dispersion, as expected from the enclosed mass. 
        Symbols are as indicated in Table \ref{tab:properties}, with dots representing the {\sc m}o{\sc ria} galaxies without resolved rotation curves.  \label{fig:dispersion}}
\end{figure}

In accordance with the analysis of P16, we have used \ion{H}{\sc i} ordered motions to get an idea of the underlying gravitational potential. Late-type dwarf galaxies are typically dispersion-supported \citep{kirby14, wheeler17}. So the stellar velocity dispersion $\sigma_\star$ of a late-type dwarf can also be used to obtain a mass estimate, by using
\begin{equation}
M_{\sigma_\star}(R) = 3 \sigma_\star^2 R G^{-1}.
\end{equation}
For our simulations, we want to calculate $\sigma_\star$ in the same way as is done observationally \citep{kirby14}. In the same vein as \citet{vandenbroucke16}, we weigh the average with the number of red giant branch (RGB) stars expected in each stellar particle \citep[using the stellar evolution models of][]{bertelli08, bertelli09}. As shown in Fig. \ref{fig:dispersion}a, the enclosed mass within $R_\mathrm{out}$ inferred from the stellar velocity dispersion agrees reasonably well with the one inferred from the \ion{H}{\sc i} kinematics ($M_\mathrm{circ, \ion{H}{\sc i}} = v_\mathrm{out, \ion{H}{\sc i}}^2 R_\mathrm{out} G^{-1}$). The mass estimated from stellar velocity dispersions agrees typically within $\sim30\%$ with the mass inferred from the HI rotation curve. Two simulations however have a relative difference of $\sim50\%$.

Fig. \ref{fig:dispersion}b shows the measured stellar velocity dispersion within $R_\mathrm{out}$ as a function of the dynamical one, $\sigma_\mathrm{dyn}$, as expected from the enclosed mass. We note that we have only included simulations with at least 100 stellar particles within $R_\mathrm{out}$, to get a good measure of $\sigma_\star$. Contrary to Fig. \ref{fig:expected_vc}, the lowest-mass \textsc{m}o\textsc{ria} dwarfs do not systematically have lower observed stellar velocity dispersions than dynamical ones. However, over our entire sample, the majority of the simulated dwarfs have $\sigma_\star < \sigma_\mathrm{dyn}$. This implies that in most cases, their dynamical mass would be underestimated from observed stellar kinematics.

Given this, the results presented in this paper might also be extended to the TBTF problem for satellite galaxies. However, other effects play an important role. The presence of \ion{H}{\sc i} and active star formation (and thus stellar feedback) in field dwarfs will influence the stellar kinematics through dynamical heating or cooling. Satellites are devoid of \ion{H}{\sc i} but will, on the other hand, be influenced by the tidal field from their host galaxy. A similar study of simulated satellite galaxies is thus necessary to see if their stellar kinematics underestimate the halo mass.

\subsection{Are disturbed velocity fields realistic?} 
As can be seen from Figs. \ref{fig:overview_sims_appendix1}-\ref{fig:overview_sims_appendix9}, 
the \ion{H}{i} distributions of the {\sc m}o{\sc ria} galaxies are generally quite disturbed. 
This is also the case for most real dwarf galaxies with velocity widths $W_{50}/2 \lesssim 30\ 
\mathrm{km~s^{-1}}$; see e.g. Leo P \citep{bernsteincooper14}, CVndwA, DDO 210, and 
DDO 216 \citep{oh15}. 
However, galaxies with larger velocity widths, $W_{50}/2 \sim 40-70 \mathrm{km~s^{-1}}$, typically display regular velocity
fields and low \ion{H}{i} velocity dispersions, $\sigma_\mathrm{\ion{H}{i}} \lesssim 12 \mathrm{km~s^{-1}}$ \citep[e.g.][]{kirby12, iorio17}. In contrast, the
most massive {\sc m}o{\sc ria} dwarf that we have analysed, M-10, has a fairly disturbed velocity field
and a relatively large velocity dispersion, $\sigma_\mathrm{\ion{H}{i}} \gtrsim 20 \mathrm{km~s^{-1}}$ (see Fig. \ref{fig:overview_sims_appendix9}). Even though we cannot draw reliable conclusions from this one object alone, it is possible that this indicates that the efficiency
of stellar feedback in the {\sc m}o{\sc ria} simulation is too strong. We note that most state-of-the-art hydrodynamic
simulations of dwarf galaxy formation \citep[e.g.][]{hopkins14, wang15, sawala16b} have more efficient feedback
schemes than {\sc m}o{\sc ria}, so this could represent a general issue for (dwarf) galaxy simulations.
At the same time however, the {\sc m}o{\sc ria} simulation succesfully reproduces the sizes and thicknesses of \ion{H}{i}
disks of observed dwarfs (Figs. \ref{fig:axisratio}  \ref{fig:sizes}). Moreover, in Fig. 9 of V15 we show that the spatial
distribution of \ion{H}{i} in {\sc m}o{\sc ria} dwarfs has similar power spectrum slopes as those measured for LITTLE
THINGS galaxies \citep{zhang12, hunter12}.
We leave the investigation of this, including, for example, the effect of beam size, for future research.

In any case, this does not 
change the conclusions of this paper in any way, since these are based on the galaxies with 
$W_{50}/2 \lesssim 30\ \mathrm{km~s^{-1}}$.

\section{Conclusions}
\label{sec:conclusions}

We have used the {\sc m}o{\sc ria} simulations of dwarf galaxies with realistic
\ion{H}{i} distributions and kinematics to investigate the
Too Big To Fail problem for late-type field dwarfs.

We showed that the {\sc m}o{\sc ria} dwarfs follow the relation between \ion{H}{i}
line-width and halo circular velocity, derived by
\cite{papastergis16}, which is required for the $\Lambda$CDM halo velocity
function to correspond to the observed field galaxy width
function. This means that, given the number density of halos formed
in a $\Lambda$CDM universe, the {\sc m}o{\sc ria} simulations reproduce the observed galactic
velocity function. In other words:~there are no missing dwarfs 
in the {\sc m}o{\sc ria} simulations.

We then constructed resolved \ion{H}{i} rotation curves, including corrections for
pressure support, for ten of the {\sc m}o{\sc ria} dwarf galaxies. We used our mock \ion{H}{i} rotation
curves to replicate the analysis of \cite{papastergis16} and fitted NFW and DC14 density profiles
(with fixed concentration parameter) to the outermost point of these
measured rotation curves to derive an observational estimate for the
maximum halo circular velocity of each {\sc m}o{\sc ria} galaxy. Using this
estimate for the circular velocity, the {\sc m}o{\sc ria} dwarfs, like the real
dwarf galaxies analysed by \cite{papastergis16}, fail to adhere to the
relation between \ion{H}{i} line-width and halo circular velocity that
is required for the $\Lambda$CDM halo velocity function to correspond to the
observed field galaxy width function. In other words,~using only
quantities derived from observations, dwarf galaxies (both real and
simulated) experience the TBTF problem.
What causes this difference between the results from fitting a halo
profile to the outer-most point of the rotation curve and using the
actual $v_{h, \mathrm{max}}^\mathrm{true}$-value derived directly from the mass
distribution? 

Comparing the \ion{H}{i} rotation curves of the {\sc m}o{\sc ria} dwarf
galaxies with their theoretical halo circular velocity curves, we see
that they can differ significantly. The circular velocities derived
from the \ion{H}{i} kinematics of {\sc m}o{\sc ria} dwarfs with \ion{H}{i}
rotation velocities below $\sim 30\ \mathrm{km\ s^{-1}}$ are typically
too low. This results in a $v_{h, \mathrm{max}}^\mathrm{fit}$ value that is too low at a
fixed concentration $c$. The TBTF problem thus results, at least 
partially, from the fact that for galaxies in this regime, their
halo mass cannot readily be inferred from their (\ion{H}{i}) kinematics. Indeed, 
based on their kinematics, galaxies with $W_{50}/2 \lesssim 30\ \mathrm{km\ s^{-1}}$ 
are predicted to inhabit halos that are less massive than observations would suggest.
However, under the assumptions of $\Lambda$CDM, the P16 relation does provide an estimate
of the true halo circular velocity $v_{h, \mathrm{max}}^\mathrm{true}$ as a function
of a galaxy's \ion{H}{i} linewidth $W_{50}$.

We attribute this effect to that fact that the atomic interstellar medium of
low-mass dwarfs simply does not form a relatively flat, dynamically
cold disc whose kinematics directly trace the underlying
gravitational force field. Another explanation might be that the \ion{H}{i} velocity 
fields are too irregular to infer halo mass from kinematics. This is true
for most of the simulated galaxies presented in this work, as well as for observed low-mass 
galaxies.
 
The stellar feedback efficiency will influence both the \ion{H}{i}-thickness of the
galaxies, as well as how messy the velocity fields are. Thus, how much energy 
is actually injected in the ISM by stellar feedback is an important issue in the
discussion of the TBTF problem.

\section*{Acknowledgements}

We thank the anonymous referee for their constructive comments, which improved the content and presentation of the paper. 
This research has been funded by the Interuniversity
Attraction Poles Programme initiated by the Belgian Science Policy
Office (IAP P7/08 CHARM). EP is supported by a NOVA postdoctoral fellowship
 of the Netherlands Research School for Astronomy (NOVA).
SDR thanks the Ghent University
Special Research Fund (BOF) for financial support. We also thank Flor
Allaert, Antonino Marasco, Arianna Di Cintio, Aurel Schneider, and Sebastian Trujillo-Gomez
for useful comments and discussions. We thank Volker Springel for making the {\sc Gadget-2} 
simulation code publicly available. 

\bibliographystyle{aa}
\bibliography{bibliography}

\begin{appendix}

\section{H\,{\sc i} catalogue}
\label{sec:catalogue}

A synthetic observation of one our simulations was already presented in Fig. \ref{fig:overview_sim}. 
Here, the rest of the {\sc m}o{\sc ria} simulations discussed in this paper are shown.
\begin{figure*}
\begin{minipage}{\textwidth}
\begin{center}
\includegraphics[width=0.9\textwidth]{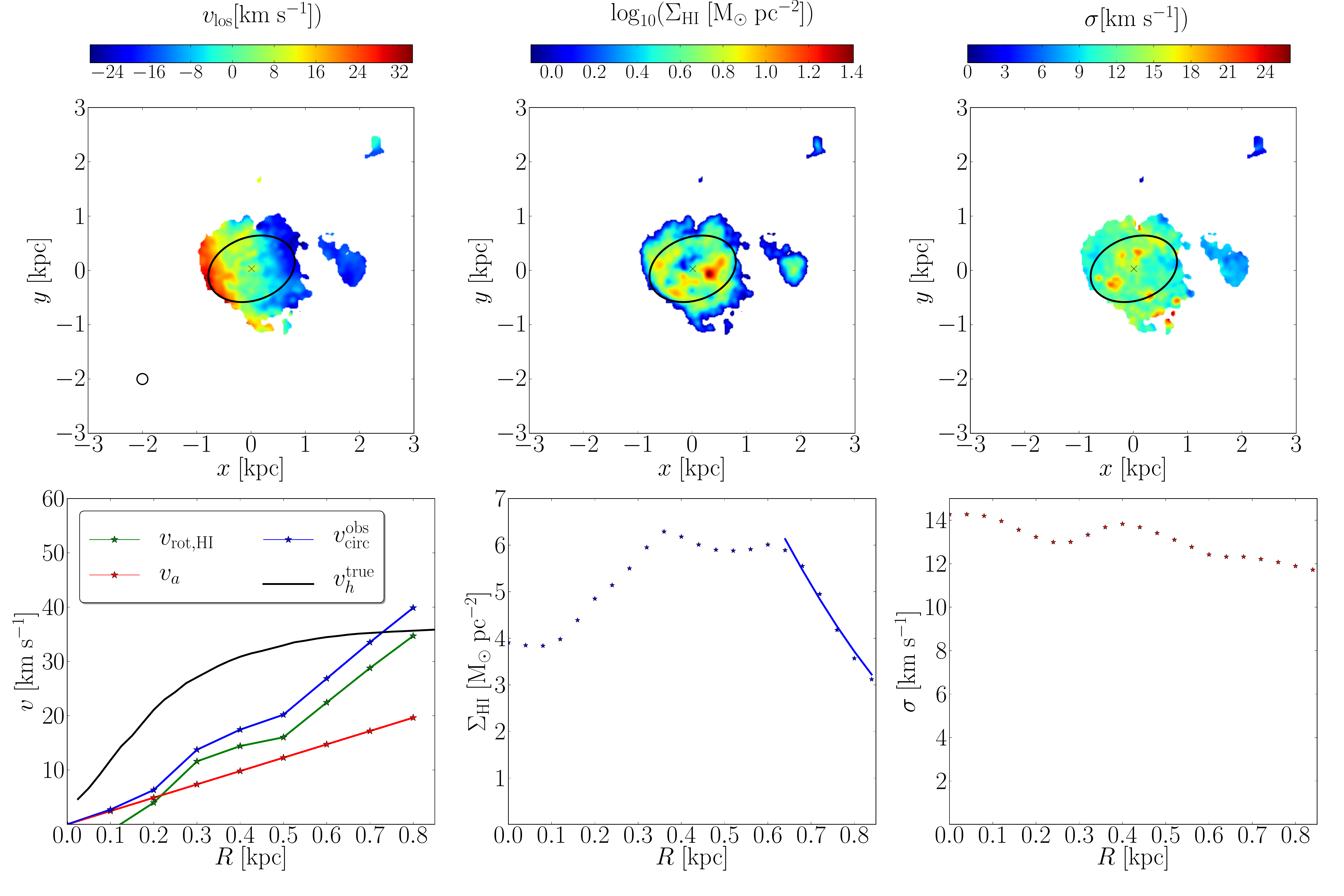}
\includegraphics[width=0.9\textwidth]{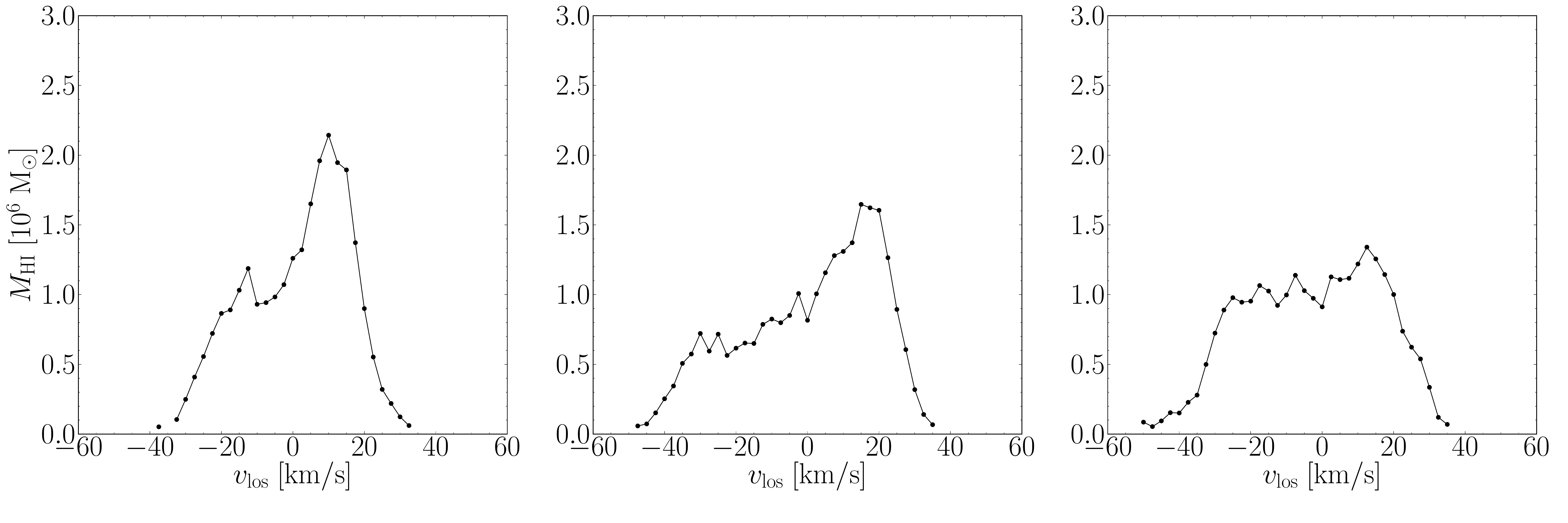}
\end{center}
\end{minipage}
\caption{Same as in Fig. \ref{fig:overview_sim}, but for M-2.}
\label{fig:overview_sims_appendix1}
\end{figure*}

\begin{figure*}
\begin{minipage}{\textwidth}
\begin{center}
\includegraphics[width=0.9\textwidth]{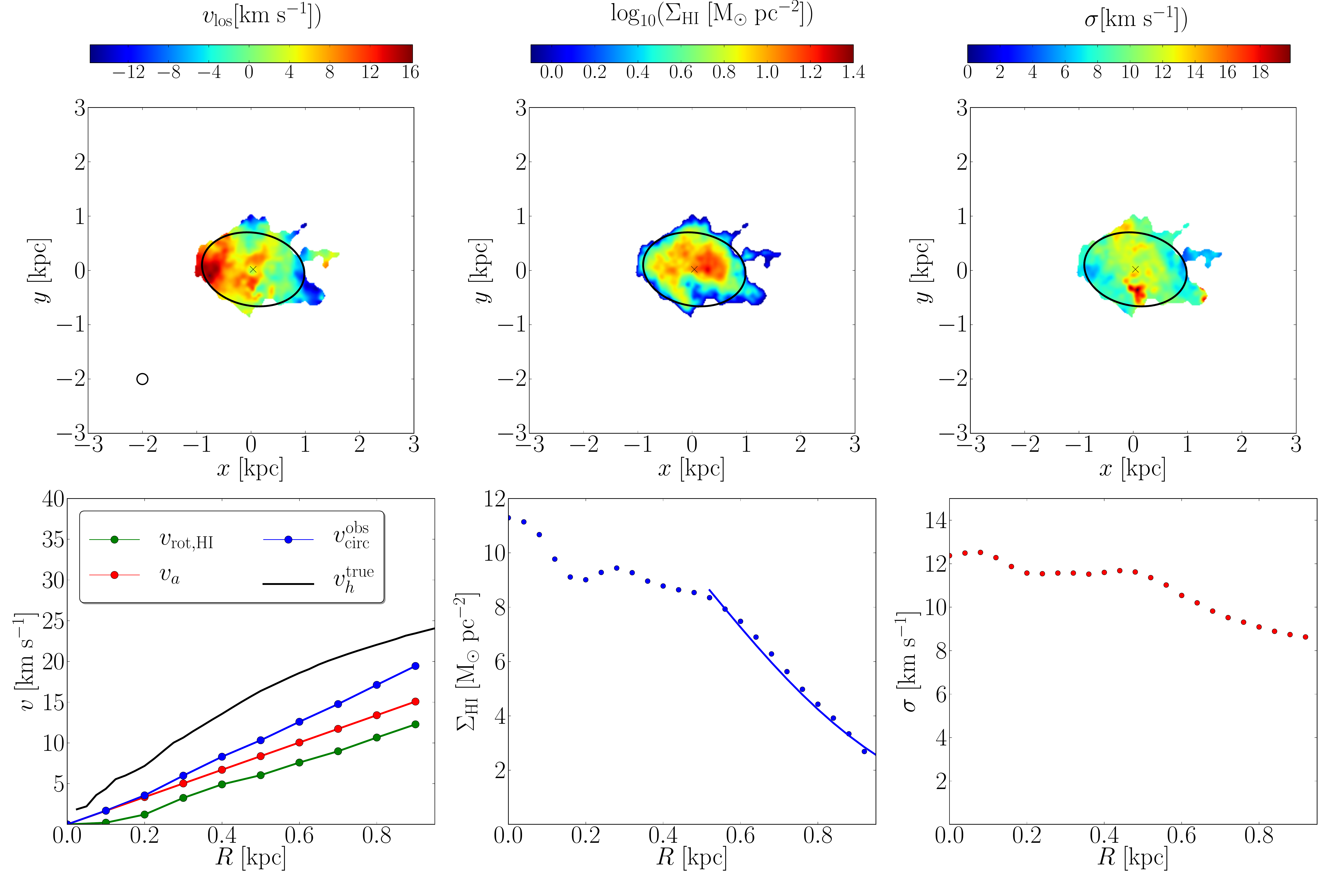}
\includegraphics[width=0.9\textwidth]{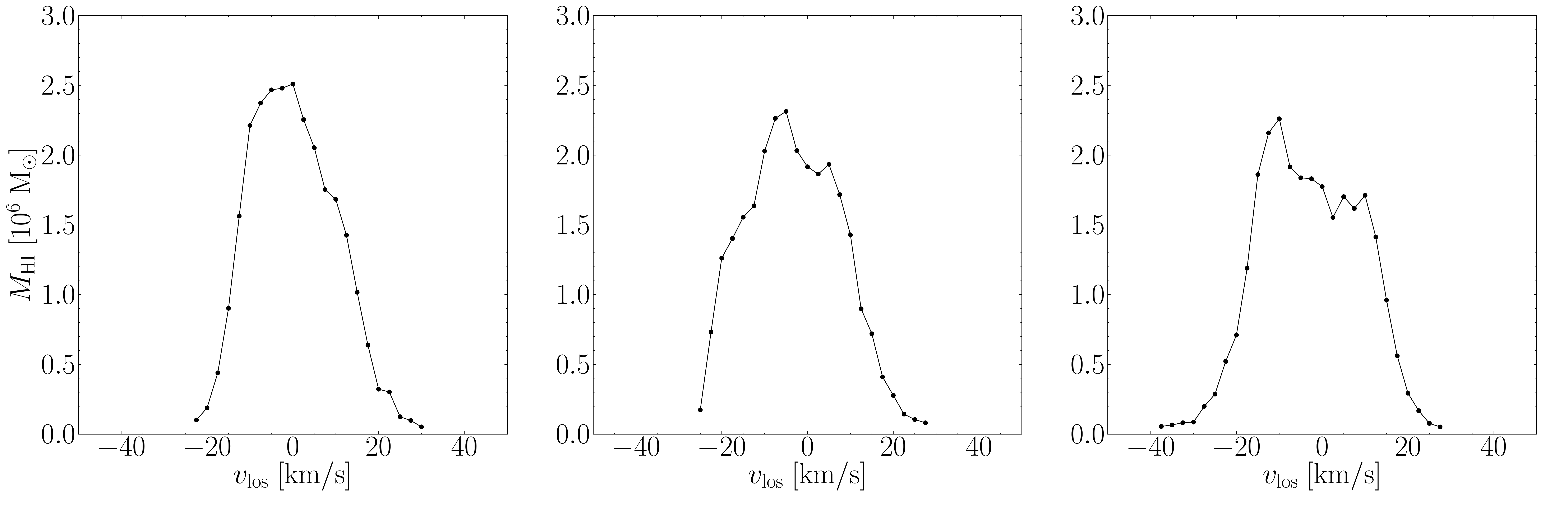}
\end{center}
\end{minipage}
\caption{Same as in Fig. \ref{fig:overview_sim}, but for M-3.}
\label{fig:overview_sims_appendix2}
\end{figure*}

\begin{figure*}
\begin{minipage}{\textwidth}
\begin{center}
\includegraphics[width=0.95\textwidth]{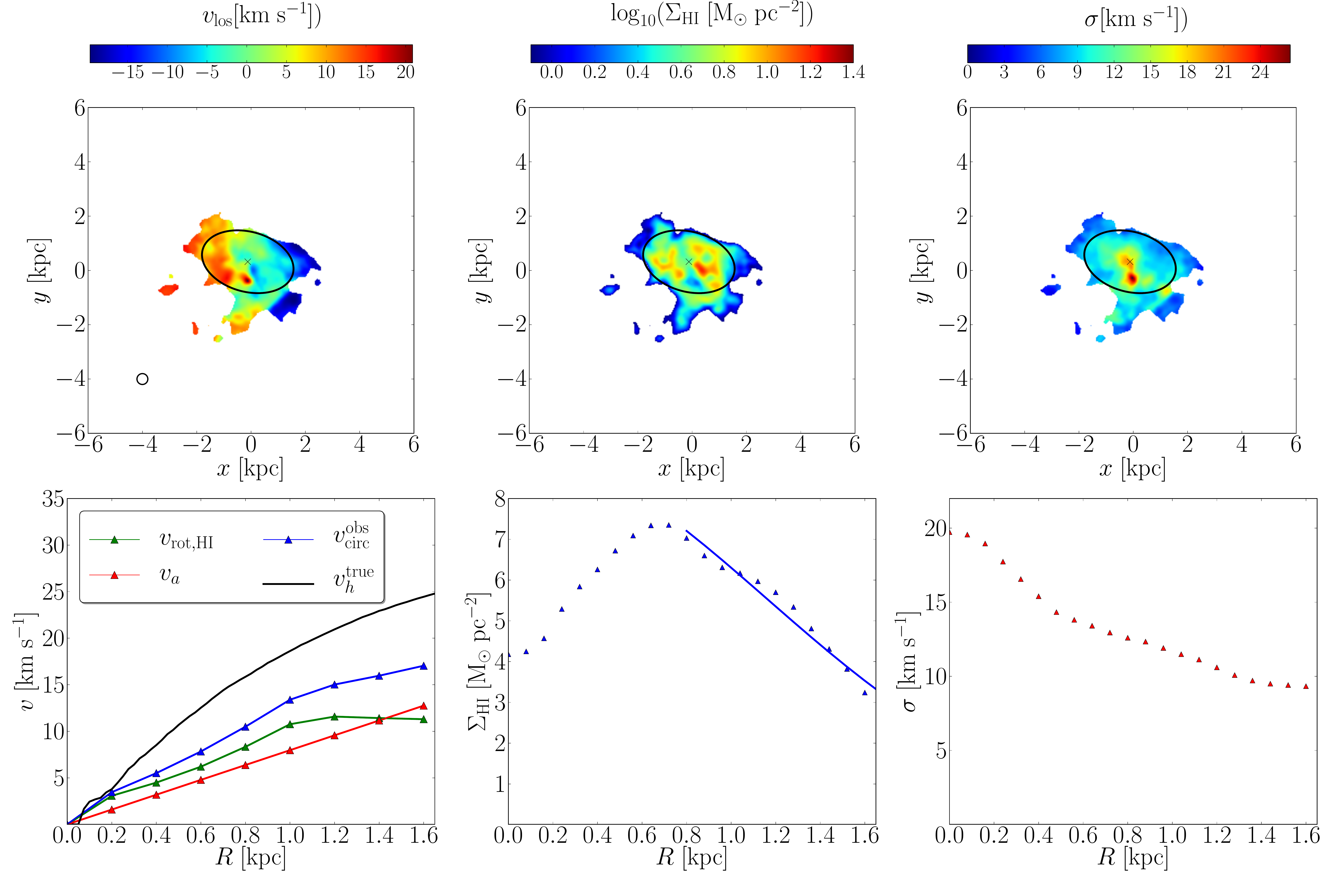}
\includegraphics[width=0.95\textwidth]{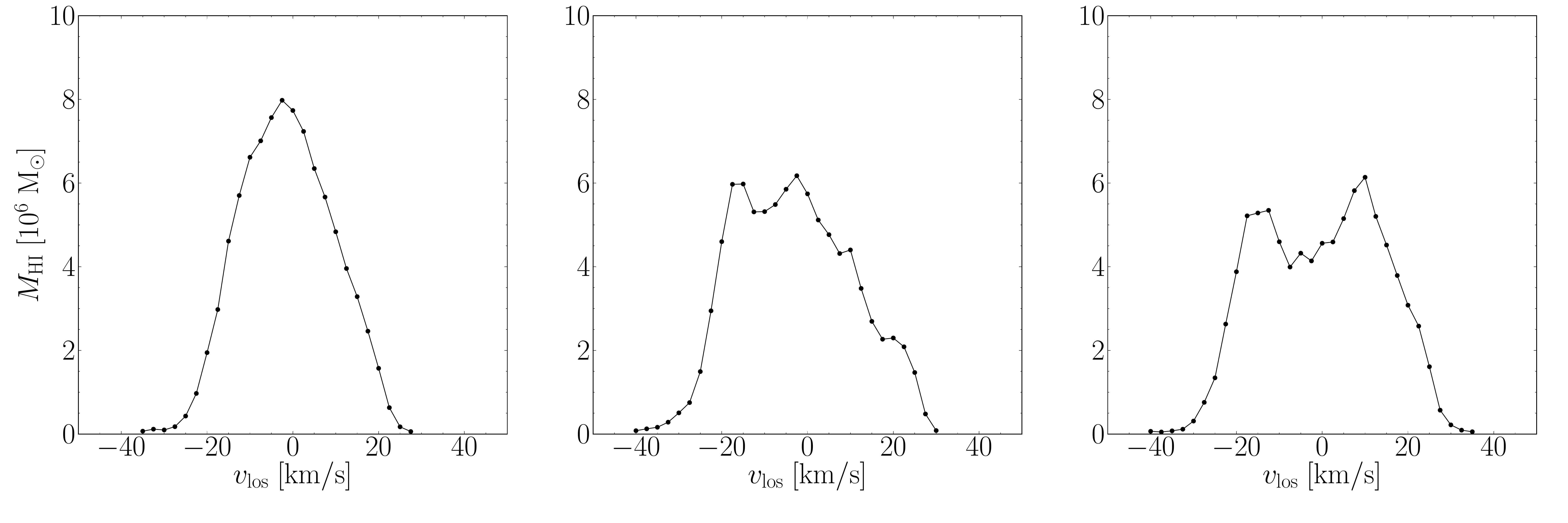}
\end{center}
\end{minipage}
\caption{Same as in Fig. \ref{fig:overview_sim}, but for M-4.}
\label{fig:overview_sims_appendix3}
\end{figure*}

\begin{figure*}
\begin{minipage}{\textwidth}
\begin{center}
\includegraphics[width=0.95\textwidth]{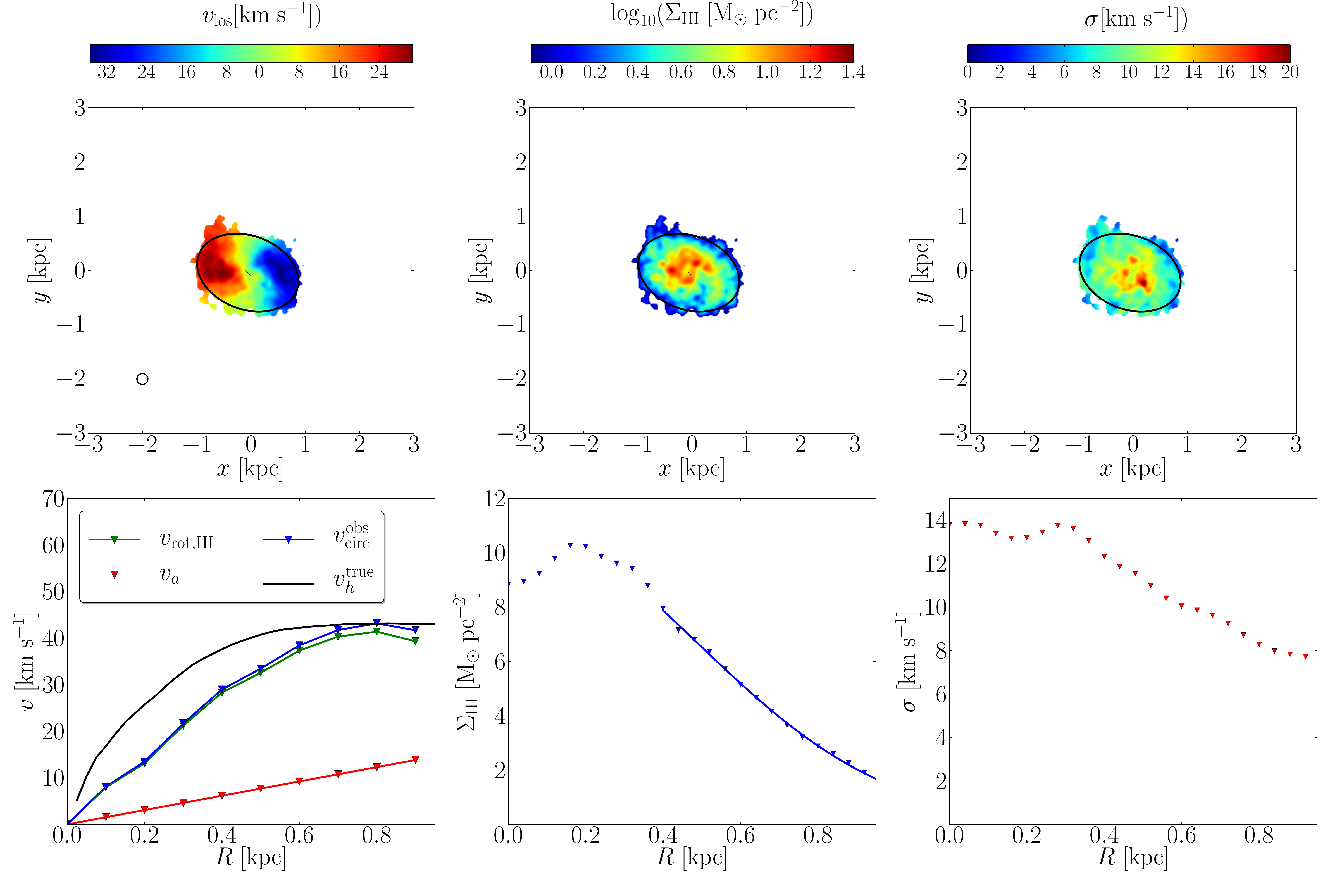}
\includegraphics[width=0.95\textwidth]{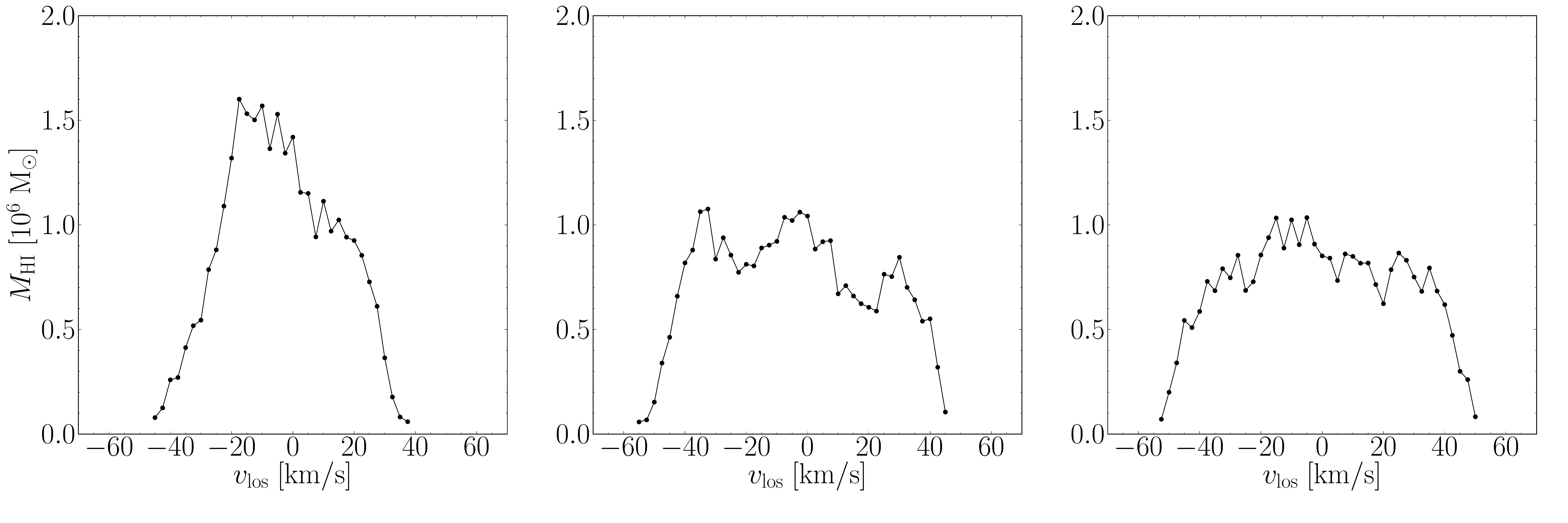}
\end{center}
\end{minipage}
\caption{Same as in Fig. \ref{fig:overview_sim}, but for M-5.}
\label{fig:overview_sims_appendix4}
\end{figure*}

\begin{figure*}
\begin{minipage}{\textwidth}
\begin{center}
\includegraphics[width=0.95\textwidth]{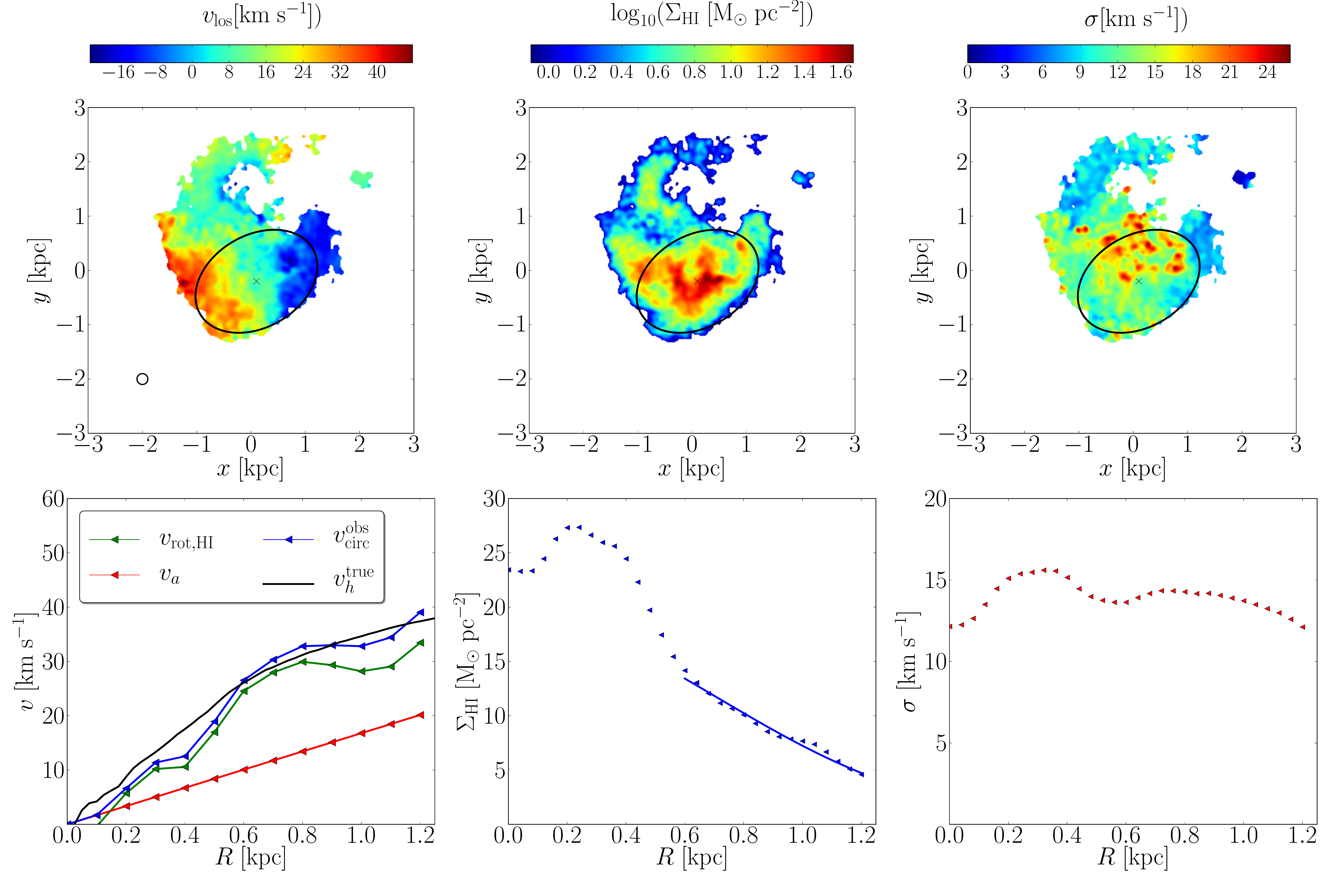}
\includegraphics[width=0.95\textwidth]{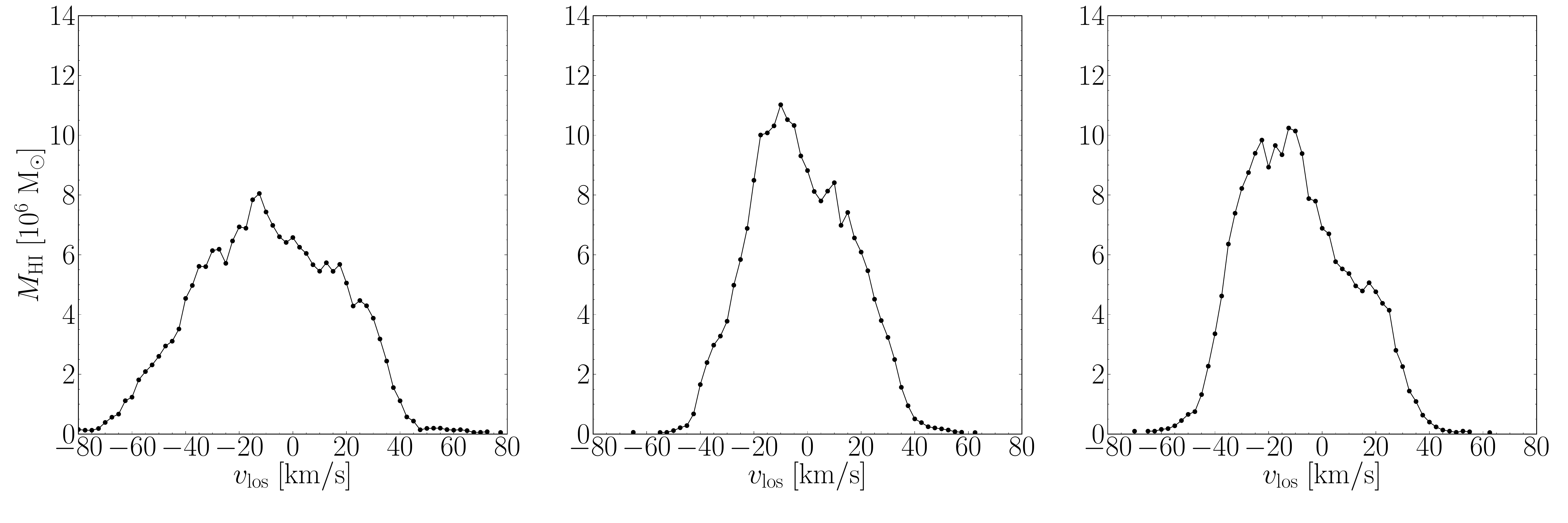}
\end{center}
\end{minipage}
\caption{Same as in Fig. \ref{fig:overview_sim}, but for M-6.}
\label{fig:overview_sims_appendix5}
\end{figure*}

\begin{figure*}
\begin{minipage}{\textwidth}
\begin{center}
\includegraphics[width=0.95\textwidth]{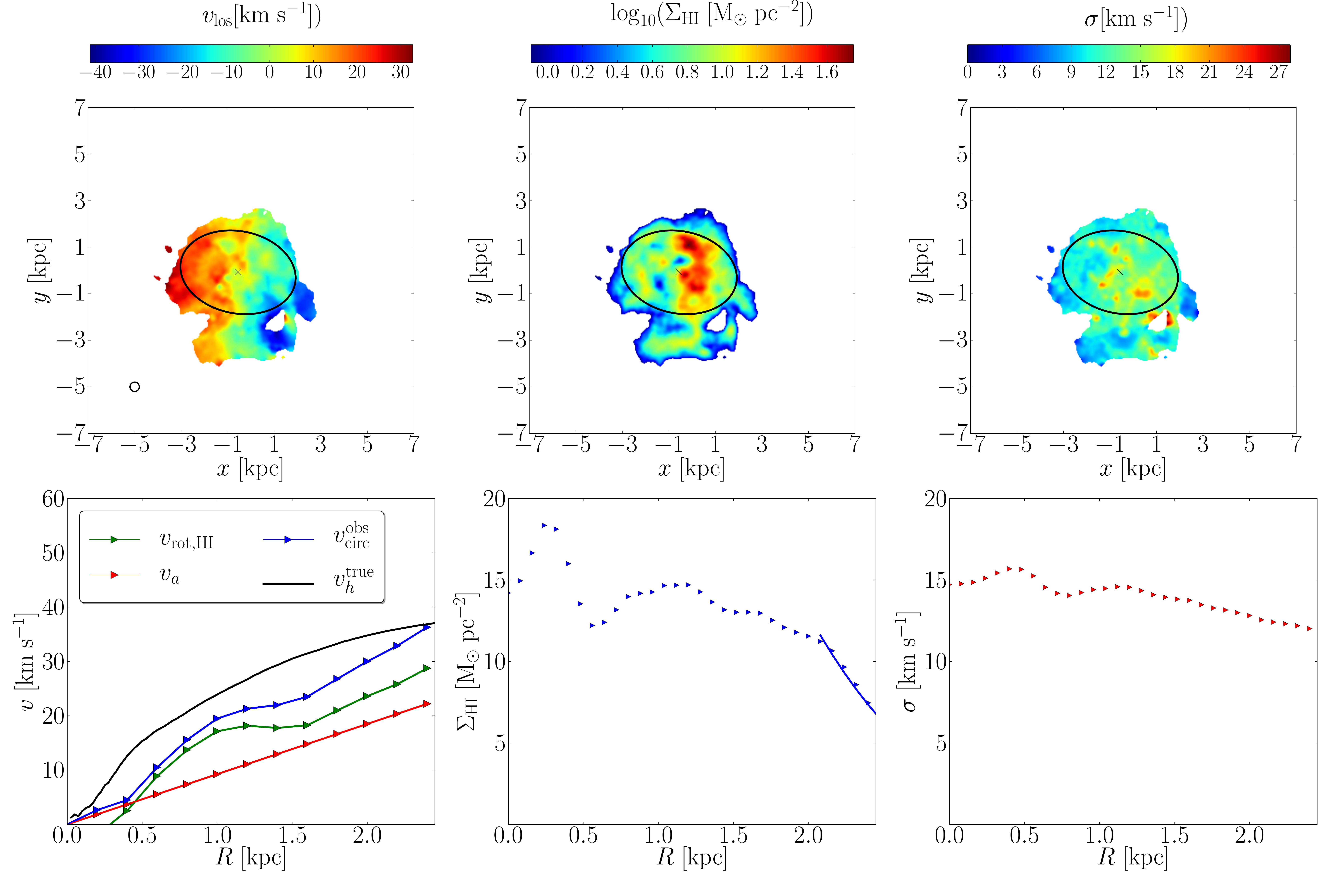}
\includegraphics[width=0.95\textwidth]{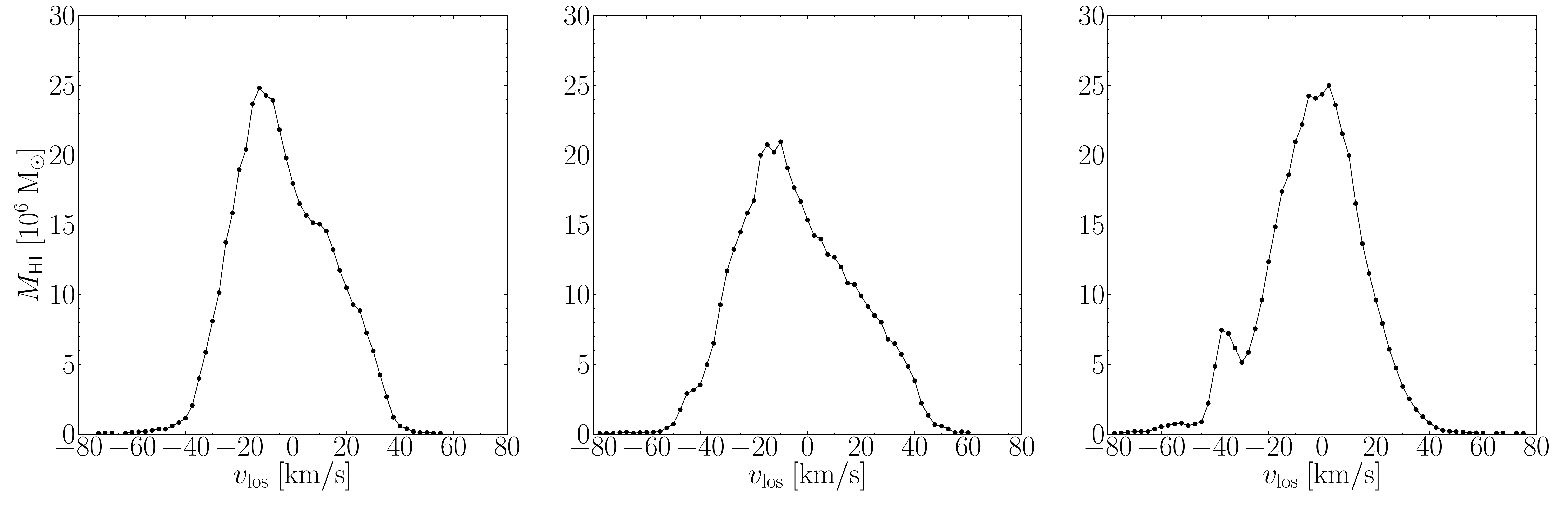}
\end{center}
\end{minipage}
\caption{Same as in Fig. \ref{fig:overview_sim}, but for M-7.}
\label{fig:overview_sims_appendix6}
\end{figure*}

\begin{figure*}
\begin{minipage}{\textwidth}
\begin{center}
\includegraphics[width=0.95\textwidth]{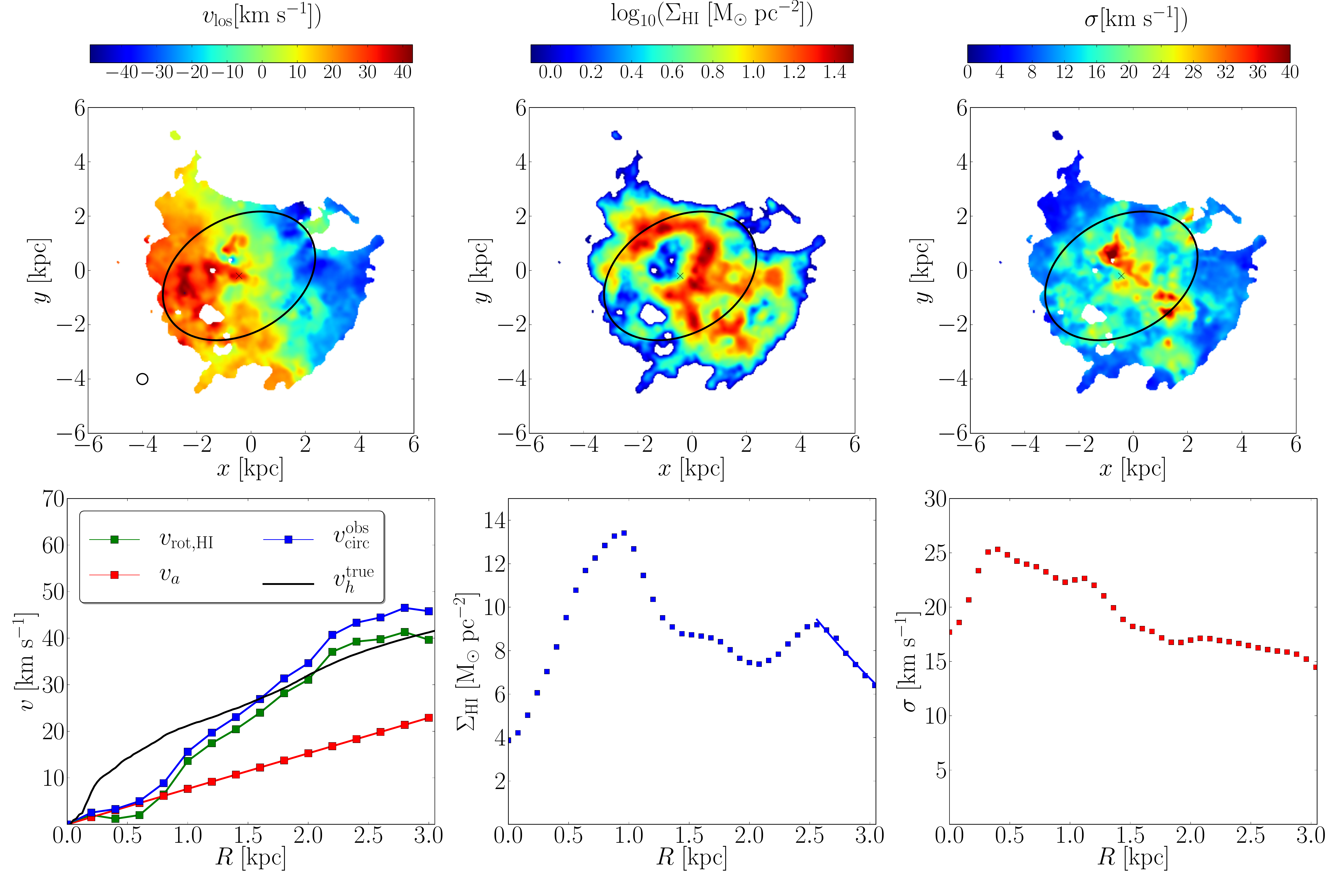}
\includegraphics[width=0.95\textwidth]{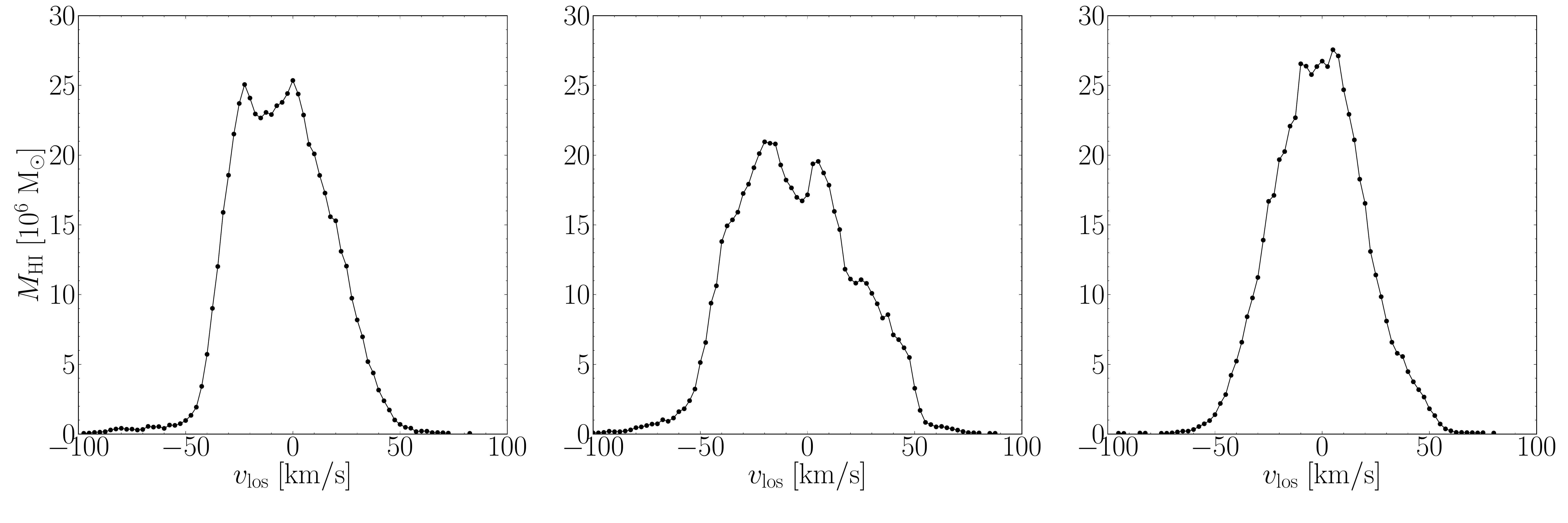}
\end{center}
\end{minipage}
\caption{Same as in Fig. \ref{fig:overview_sim}, but for M-8.}
\label{fig:overview_sims_appendix7}
\end{figure*}

\begin{figure*}
\begin{minipage}{\textwidth}
\begin{center}
\includegraphics[width=0.95\textwidth]{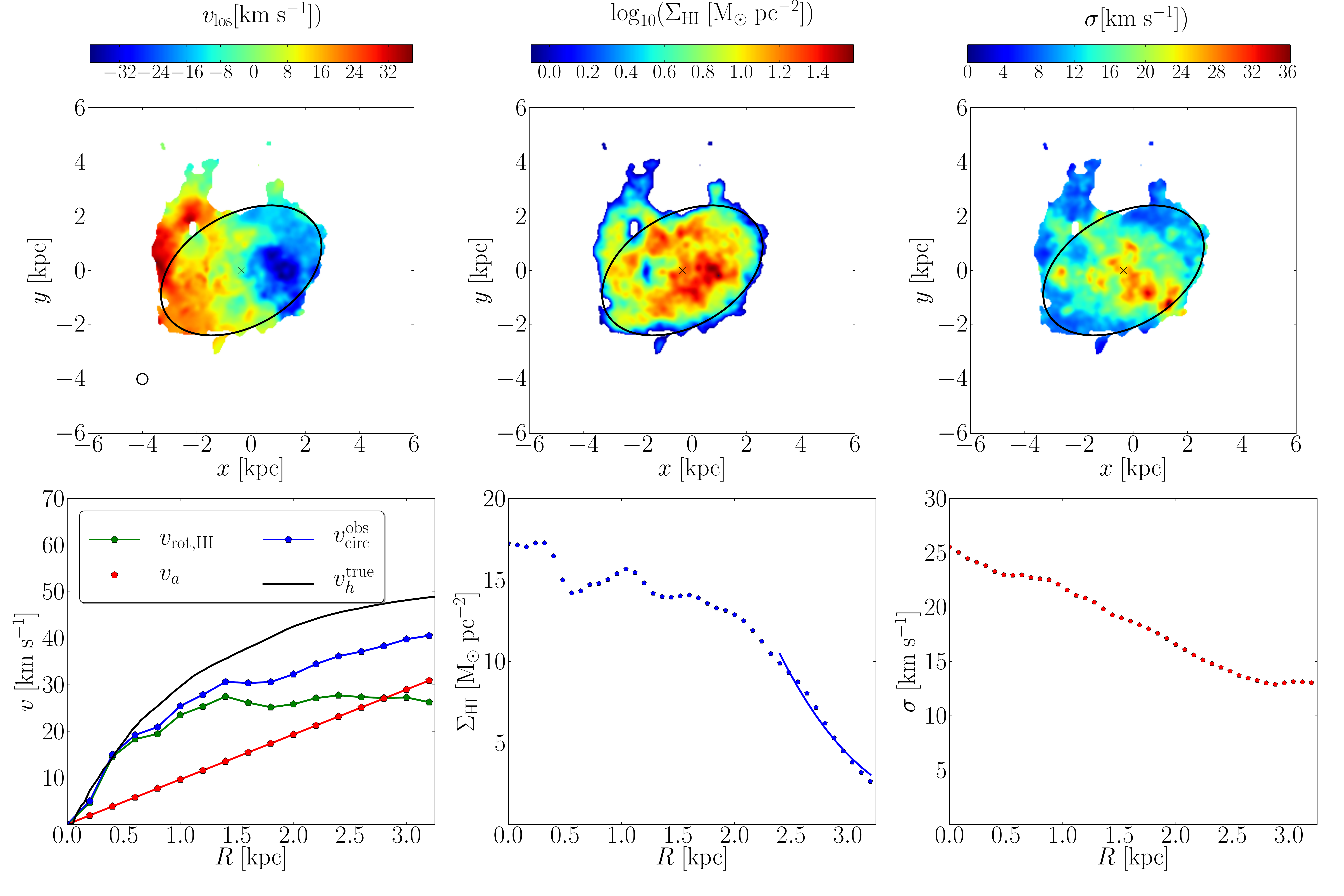}
\includegraphics[width=0.95\textwidth]{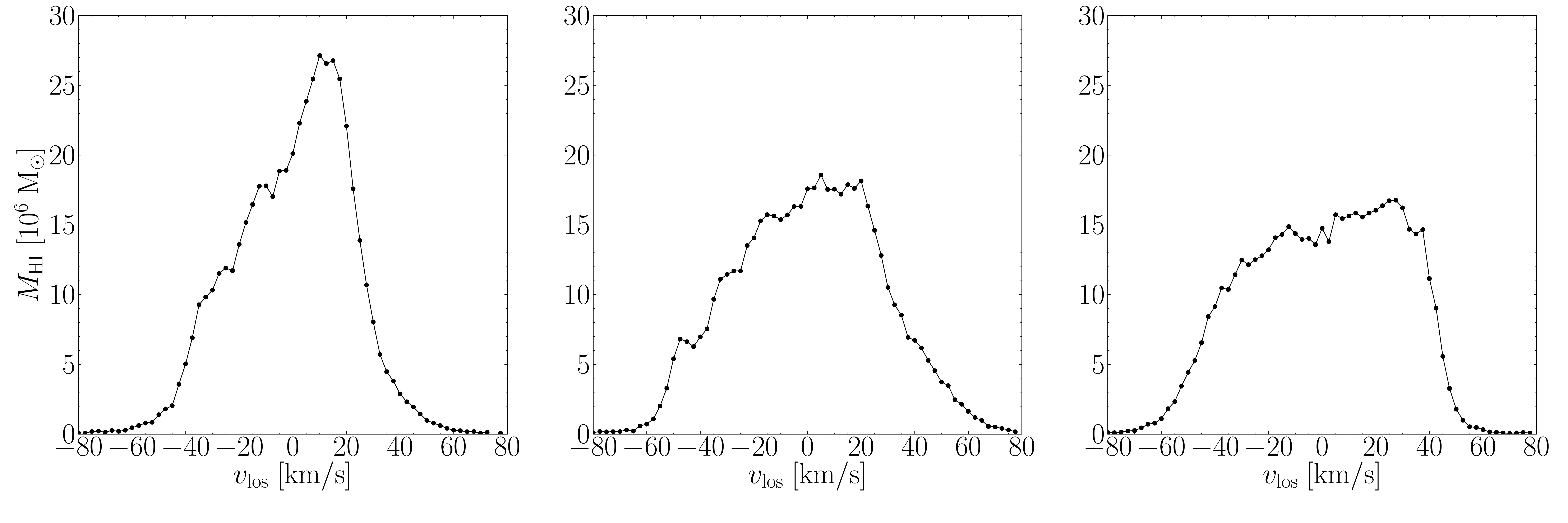}
\end{center}
\end{minipage}
\caption{Same as in Fig. \ref{fig:overview_sim}, but for M-9.}
\label{fig:overview_sims_appendix8}
\end{figure*}

\begin{figure*}
\begin{minipage}{\textwidth}
\begin{center}
\includegraphics[width=0.95\textwidth]{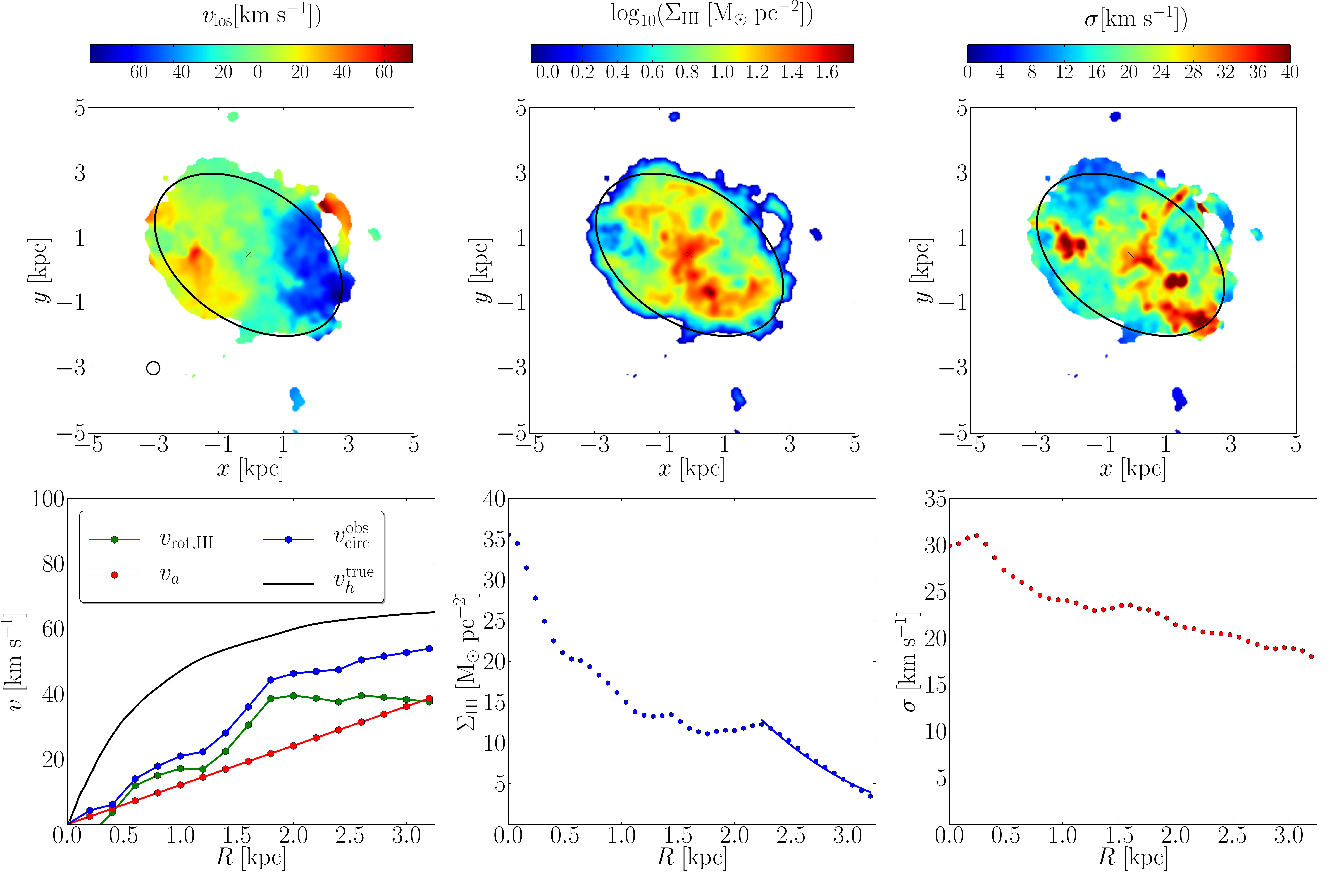}
\includegraphics[width=0.95\textwidth]{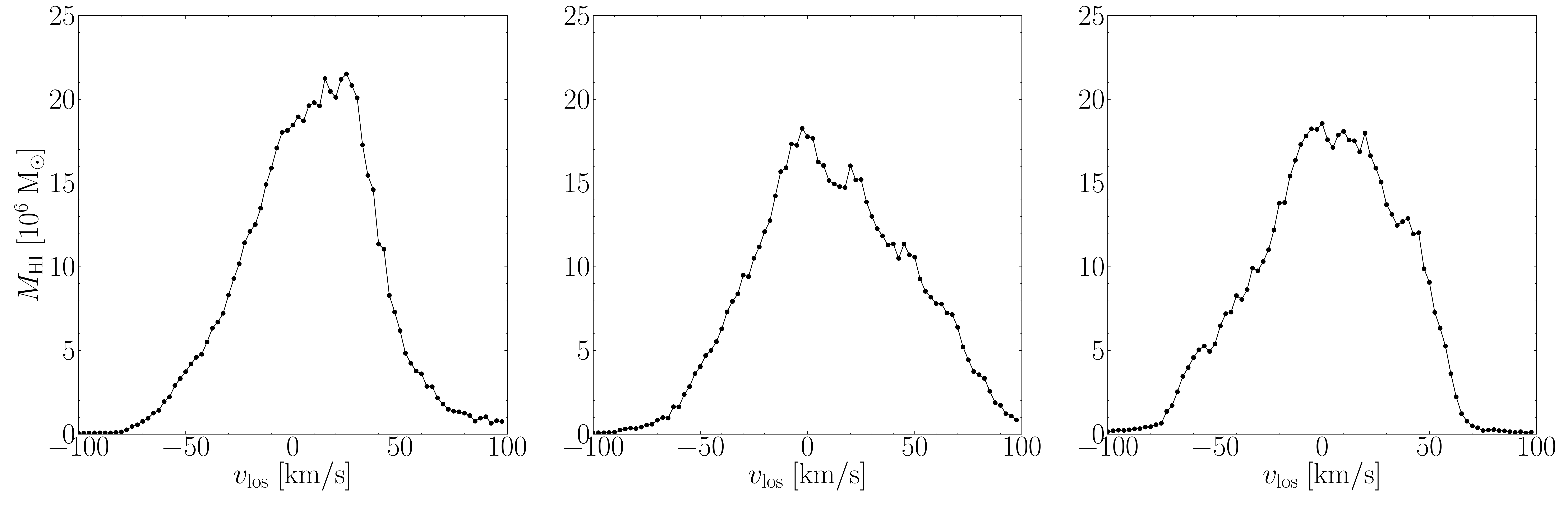}
\end{center}
\end{minipage}
\caption{Same as in Fig. \ref{fig:overview_sim}, but for M-10. }
\label{fig:overview_sims_appendix9}
\end{figure*}

\section{Concentration fitting}
\label{sec:concentrationfitting}

To be able to fit the two-parameter density profiles given by
Eqs. (\ref{eq:nfw}) and (\ref{eq:dc14}) to only two points (the
central and outermost measured point of the rotation curve), P16 kept
the concentration $c$ of the halos fixed at the mean cosmic value and
used the halo mass $M_h$ as a free parameter. Another choice would be
to fix $M_h$ using the stellar mass and an abundance matching relation
and to keep the concentration $c$ as a free parameter.

\begin{figure}
\includegraphics[width=0.47\textwidth]{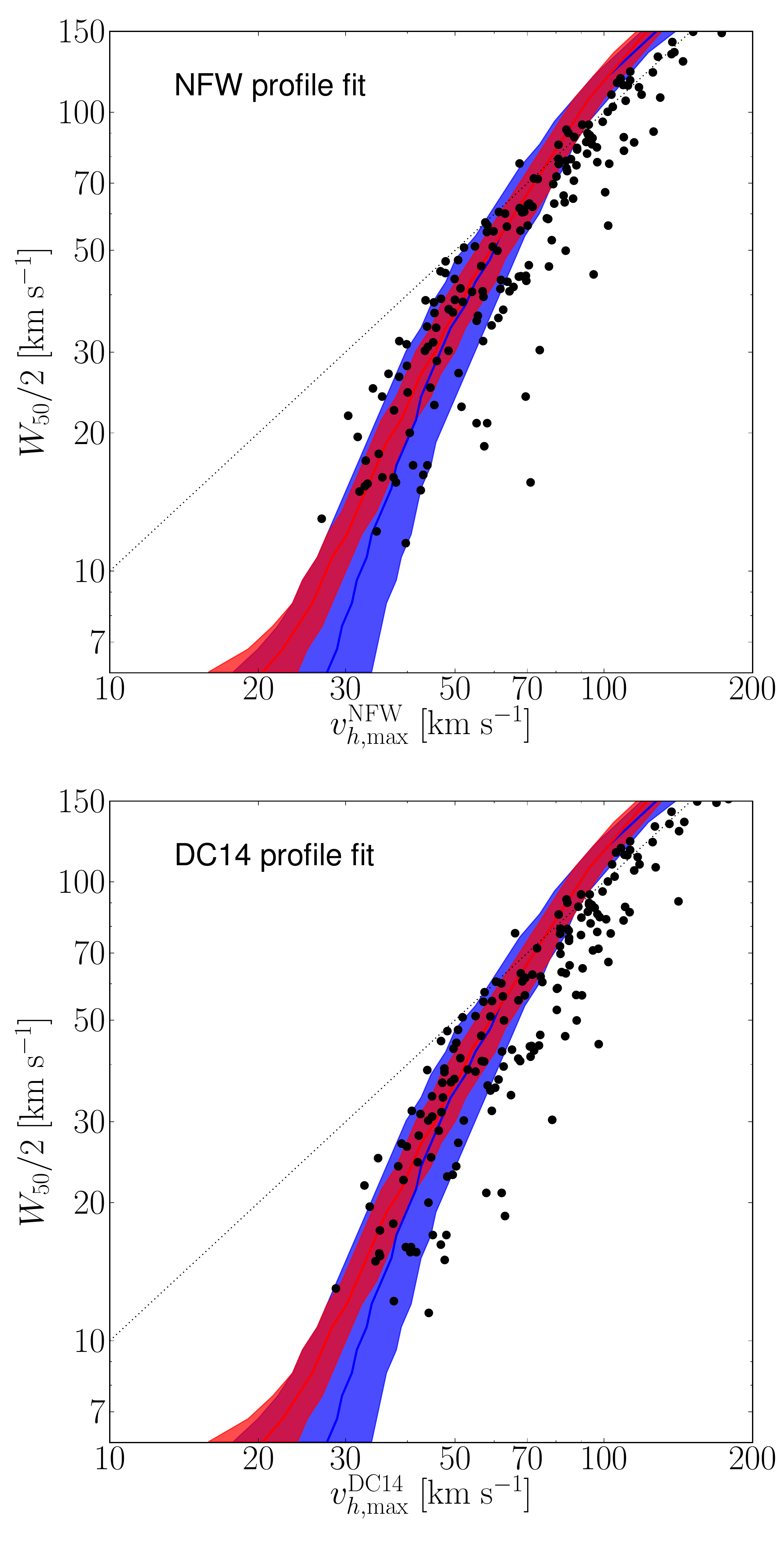}
\caption{Results from fitting an NFW (top panel) and DC14 (bottom
  panel) to the outer-most point of the rotation curves of the
  observations used in \citet{papastergis16} using a fixed halo mass,
  keeping the halo concentration as a free parameter. The halo mass is
  calculated from their stellar mass using the abundance-matching
  relation of \citet{moster13}. Red and blue lines and bands are the
  same as in Fig. \ref{fig:halofit}. \label{fig:concentration_fit}}
\end{figure}

In Fig. \ref{fig:concentration_fit}, we show the $W_{50}-v_{h,
  \mathrm{max}}^\mathrm{fit}$ relation obtained by fitting NFW and DC14 profiles to
the P16 dataset, using the concentration $c$ as a free parameter with
the halo mass $M_h$ set by the stellar mass and the \citet{moster13}
abundance matching relation. This way, the observed galaxies adhere
much more closely to the expected $W_{50}-v_{h, \mathrm{max}}^\mathrm{true}$
relation. NFW and DC14 profiles now actually produce very similar
results. 

\begin{figure}
\includegraphics[width=0.47\textwidth]{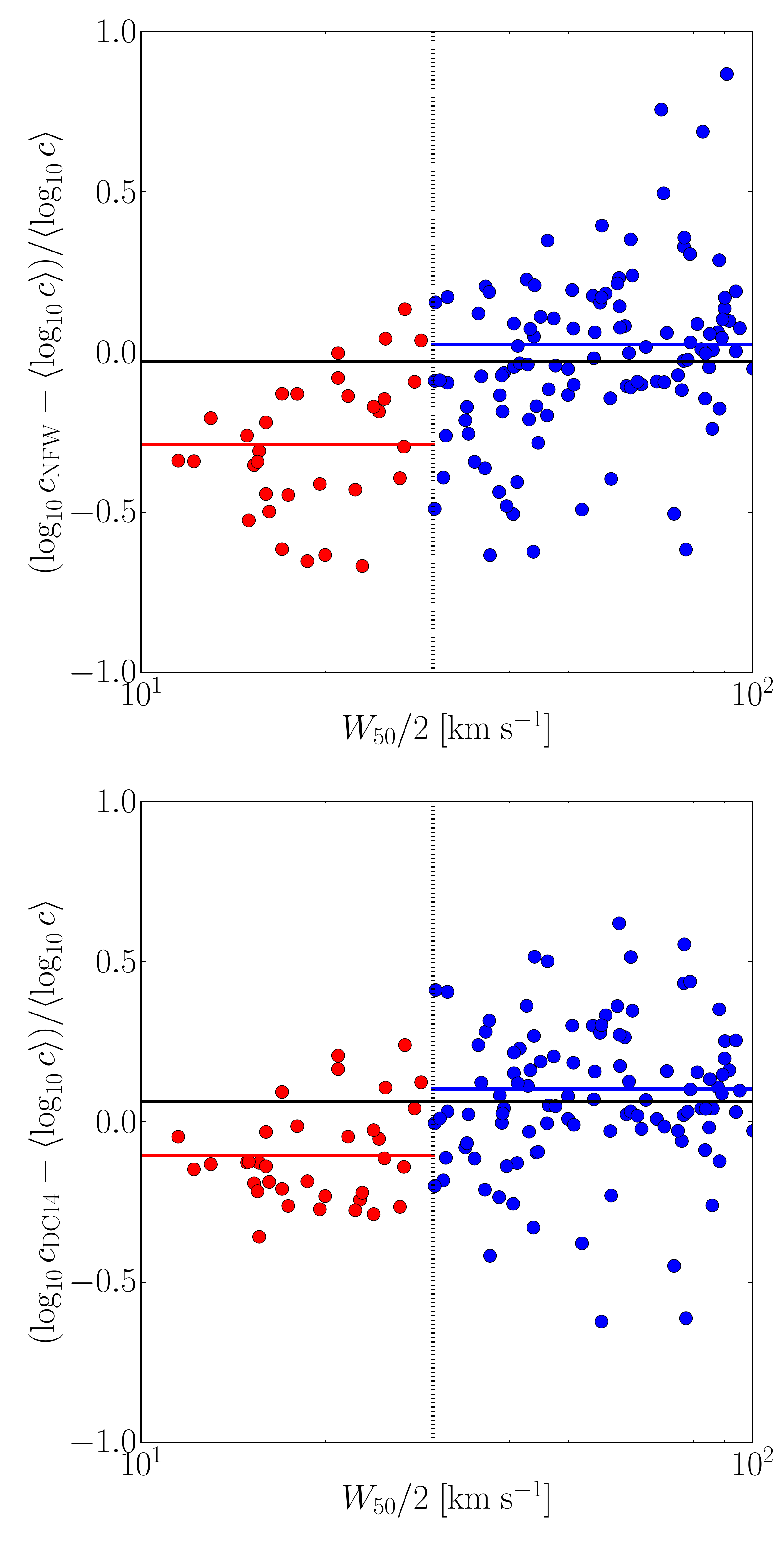}
\caption{Logarithmic difference between the fitted concentration and
  the one expected from cosmological dark-matter only simulations
  \citep{dutton14} of the P16 sample of galaxies. Top
  panel:~concentration obtained by fitting the NFW density profile to
  the kinematic data; bottom panel:~concentration obtained using the
  DC14 density profile. Red and blue symbols indicate galaxies with
  low and high \ion{H}{i} rotation velocities, respectively. The red
  and blue lines indicate the mean concentration of both subsamples;
  the green line indicates the mean concentration of the full sample.
\label{fig:concentration_diff}}
\end{figure}

In Fig. \ref{fig:concentration_diff}, we compare the concentrations of
the P16 galaxies retrieved in this way with the mass-dependent cosmic
mean value derived from cosmological simulations \citep{dutton14}. The
frequency distribution of the concentration values is well
approximated with a log-normal distribution function. Both for a NFW
and a DC14 fit, the scatter is $\sigma \approx 0.25-0.3$~dex. This is
significantly larger than the scatter on $\log(c/\langle c \rangle)$
found in cosmological simulations, where $\sigma \sim 0.13$
\citep{dutton14}. 

In Fig. \ref{fig:concentration_diff}, we distinguish between galaxies
with $W_{50}/2 < 30\ \mathrm{km\ s^{-1}}$ (red data-points) and $W_{50}/2
> 30\ \mathrm{km\ s^{-1}}$ (blue data-points) for the fits with the
NFW and DC14 profiles. We choose the $30\ \mathrm{km\ s^{-1}}$ split
because this is the rotation velocity below which the TBTF problem
becomes apparent. Clearly, the high-$W_{50}$ galaxies have higher
concentrations than expected while the low-$W_{50}$ dwarfs have lower
concentrations, with a hint of an anticorrelation between $M_h$ and
concentration for the low $W_{50}$ galaxies. This is to be
expected;~higher-mass galaxies must have lower concentrations in order
to have low circular velocities. To see whether or not the average of each
subsample differs significantly from the cosmic mean value, we ran a
$t$-test on both populations. We find a $p$-value of 
$1.1\times 10^{-8}$ for the galaxies with low
$W_{50}$ in the NFW case. The full and high $W_{50}$ samples have a mean concentration consistent with the
\citet{dutton14} simulations. For
the DC14 profile, the same trend is found, with all the averages
slightly higher than for the NFW profile. The $p$-values for the $t$-test are $3.4\times
10^{-4}$ for the entire sample and $4.3\times 10^{-7}$ and $2.2\times 10^{-4}$
for the high and low circular velocity samples, respectively. Employing
the DC14 density profile yields concentration estimates that are
inconsistent with the \citet{dutton14} simulations, both for the full
sample and the subsamples.

Both the large scatter and the offsets are probably due, at least in
part, to uncertainties on the (extrapolated) low-mass end of the
$M_\star-M_h$ relation that was used to derive the halo mass from the
stellar mass. In the mass regime we are interested in, the scatter on
the $M_\star-M_h$ relation is expected to be substantial
\citep[e.g.][]{sales17} and, using our approach, this
translates in an increased scatter on the concentration
parameter. Moreover, there is great variation among the different
published $M_\star-M_h$ relations in the regime of dwarf galaxies
($M_h \approx 10^{10} \mathrm{M_\odot}$). Simulations also show a large
scatter in stellar mass for these type of halos
(e.g. Fig. 7 in V15). We redid our analysis adopting
different $M_\star-M_h$ relations. Using the relation 
of \citet{guo10}, we reach the same conclusions as for the relation of \citet{moster13};
when using the relation of \citet{behroozi13}, the fitted concentrations
are more in line with the predictions from $\Lambda$CDM, however they do not follow
the P16-relation. 

\citet{katz17} fitted a NFW and a DC14-profile to 147 SPARC-galaxies, taken from a sample of 
175 galaxies with extended \ion{H}{i} rotation curves \citep{lelli16b}. They conclude that
the fitted halo masses and concentrations for the DC14-profile are in line with the predictions from 
$\Lambda$CDM. The difference between our analyses is that they only discuss
their entire sample, which consists mostly of high-mass galaxies, whereas our analysis has focused on low-mass 
galaxies ($M_\star \lesssim 10^8~\mathrm{M_\odot}$). They also use full rotation curves
to fit the halo profile to the galaxies, allowing them to fit the halo mass and concentration
simultaneously. 

By fitting a \emph{coreNFW} profile \citep{read16} to full rotation curves of a subset of the 
Little THINGS galaxies \citep{iorio17}, \citet{read17} find that these isolated dwarf 
galaxies inhabit halos consistent with the abundance-matching
relation of \citet{behroozi13} and, as such, do not find a TBTF for isolated galaxies {at all}. 
These conclusions would change when assuming a different $M_\star-M_\mathrm{halo}$ relation, as they remark
in their Appendix C. Even so, we still find that when using the relation of \citet{behroozi13}, the 
observations do not follow the P16-relation.

\end{appendix}

\label{lastpage}

\end{document}